\documentclass[twocolumn, prd, aps, nofootinbib,amssymb]{revtex4-1}
\usepackage{hyperref}
\usepackage{tikz} 
\usepackage{bigints}
\usepackage[mathscr]{eucal}
\usepackage{subfig}
\usepackage{graphicx}
\usepackage{dcolumn}
\usepackage{mathrsfs}
\usepackage{amsmath}
\usepackage{bm}
\usepackage{float}
\usepackage{wasysym}
\usepackage{stackengine}
\usepackage{lipsum}

\begin{document}

\preprint{APS/123-QED}

\title{How loud are echoes from Exotic Compact Objects?}

\author{Lu\'is Felipe Longo Micchi}
\email[]{luis.longo@ufabc.edu.br}
\affiliation{Center for Natural and Human Sciences, UFABC, Santo Andr\'e, SP  09210-170, Brazil}

\affiliation{
  Perimeter Institute for Theoretical Physics, 31 Caroline Street North, Waterloo, Ontario, N2L 2Y5, Canada
}
\author{Niayesh Afshordi}%
 \email{nafshordi@pitp.ca}
\affiliation{Department of Physics and Astronomy, University of Waterloo,
200 University Ave W, N2L 3G1, Waterloo, Canada
}

\affiliation{Waterloo Centre for Astrophysics, University of Waterloo, Waterloo, ON, N2L 3G1, Canada
}

\affiliation{
  Perimeter Institute for Theoretical Physics, 31 Caroline Street North, Waterloo, Ontario, N2L 2Y5, Canada
}

\author{Cecilia Chirenti}
\email[]{chirenti@umd.edu}
\affiliation{Department of Astronomy, University of Maryland, College Park, Maryland 20742, USA}
\affiliation{Astroparticle Physics Laboratory NASA/GSFC, Greenbelt, Maryland 20771, USA}
\affiliation{Center for Research and Exploration in Space Science and Technology, NASA/GSFC, Greenbelt, Maryland 20771, USA}
\affiliation{Center for Mathematics, Computation and Cognition, UFABC, Santo Andr\'e-SP, 09210-170, Brazil}

\date{\today}

\begin{abstract}

The first direct observations of gravitational waves (GWs) by the LIGO collaboration have motivated different tests of General Relativity (GR),
including the search for extra pulses following the GR waveform for the coalescence of compact objects. 
The motivation for these searches comes from the alternative proposal that the final compact object could differ from a black hole (BH) by the lack of an event horizon and a central singularity. Such objects are expected in theories that, motivated by quantum gravity modifications, predict horizonless objects as the final stage of gravitational collapse. In such a hypothetical case, this exotic compact object (ECO) will present a (partially) reflective surface at $r_{\rm ECO}=r_{+}(1+\epsilon)$, instead of an event horizon at $r_{+}$. For this class of objects, an in-falling wave will not be completely lost and will give rise to secondary pulses, to which recent literature refers as \textit{echoes}. However, the largely unknown ECO reflectivity is determinant for the amplitude of the signal, and details also depend on the initial conditions of the progenitor compact binary. Here, for the first time, we obtain estimates for the detectability of the first echo, using a perturbative description for the inspiral-merger-ringdown waveform and a physically-motivated ECO reflectivity. 
Binaries with comparable masses will have a stronger first echo, improving the chances of detection. For a case like GW150914, the detection of the first echo will require a minimum ringdown signal-to-noise ratio (SNR) in the range $\sim 20-60$. The most optimistic scenario for echo detection could already be probed by LIGO in the next years. With the expected improvements in sensitivity we estimate one or two events per year to have the required SNR for the first echo detection during O4. 
\end{abstract}

\maketitle

\section{Introduction}\label{intro}

Starting in 2016, the LIGO-Virgo collaboration has announced 15 (and counting)  gravitational wave (GW) signals from the coalescence of compact binaries \cite{LigoCatalog,Ligo0,Ligo8, Ligo9,Ligo10}. At least one independent group  has claimed additional detections in the LIGO data \cite{IASdetection}. Therefore, we find ourselves in a singular period in history: we can now directly probe the structure and the existence of event horizons for the first time. Far from being a trivial goal \cite{Abramowicz}, the proof of the (non) existence of black  hole (BH) horizons, which are predicted by General Relativity (GR), may exclusively rely on the search for non-GR signatures on GW signals. Other probes (e.g., electromagnetic signatures) can only test the space-time geometry up to, approximately, the light-ring.

Recently, the GW astronomy community has been faced with an important issue: does the observation of a BH quasinormal mode (QNM) \cite{Kokkotas1999, Berti_qnm_review} spectrum unequivocally imply the existence of an event horizon? This question, first proposed in \cite{Cardosoprobehorizon}, generated a prolific and ongoing debate  (see  \cite{AfshordiQuantumReview} for a recent work and \cite{CardosoReviewECOS} for a review). 

One may speculate that the event horizon is replaced by a (partially) reflective wall for BH mimickers, such as in firewall and fuzzball models (even though their refletivity is also a subject of controversy \cite{Afshordi2,Fuzzball}). The new boundary can trap  GWs between the angular momentum barrier and the reflective wall, creating an effective acoustic cavity in which the perturbation resonates. The presence of this resonating chamber causes a unique signature in the GWs: secondary pulses appear additionally to the QNM ringing expected from a black hole \cite{Cardosoprobehorizon, PaniAnalytical1,PaniAnalytical,Mark:2017dnq, Micchi1, Buenokerrlike, AfshordiManual}. In the recent literature, such pulses have been called \textit{echoes}.

We should understand this theoretical finding as more than a mere academic exercise. If echoes are eventually detected, indicating the nonexistence of event horizons, we would be laying the observational foundations of quantum gravity. Currently, and still in the near future, it is possible that echo signals (if they exist) will be below the LIGO/Virgo sensitivity curve. If that is the case we will be able to impose upper limits on their amplitude, i.e., to constrain the reflectivity of the wall.

 The presence of echoes for sufficiently compact \textit{exotic compact objetcs} (ECOs) seems to be a general feature. \footnote{An exception is the analysis presented in  \cite{Schwarzschildstars}, where the inner boundary condition for the scattering of axial gravitational waves in an ultra compact Schwarzschild star spacetime is placed at a null surface and no echoes are observed.} If an ECO is not almost exactly as compact as a BH, it is expected that its QNM spectrum differs from the BH's \cite{DeBenedictis,2009PhRvD..80l4047P, Chirenti2}. The analysis of the observed QNM ringing can thus make the distinction between the two objects, as was performed for the gravastar model \cite{Chirenti1}. During this work we focus on ECOs that are sufficiently compact and reflective to  emit distinct echoes.

Further investigation of the physics underlying  echo waveforms is needed if one wishes to properly perform an echo search. Our understanding of echoes has progressed in different fronts. The scalar case served as a first toy model to understand different phenomena related to echoes, as in \cite{Buenokerrlike, Mark:2017dnq, Micchi1}. Some works made efforts to describe the effect of the orbital motion on echo waveforms \cite{Mark:2017dnq, Micchi1,sago2020gravitational}. However, their results are mostly restricted to the use of a plunging orbit from the innermost stable circular orbit (ISCO). Investigations about the influence of a non-zero rotation of the ECO were  performed, leading to the description of phenomena as the ergoregion instability \cite{Paniquench}, mode mixing \cite{PaniAnalytical} and beating of echoes \cite{Micchi1, Holdomnewwindows}. Most of the investigations on the gravitational case were restricted to the use of gaussian-like packages as initial condition \cite{Holdomnewwindows, HoldomWaves, PaniAnalytical, AfshordiSeismology, AfshordiSeismologyII}. Although working with a more physically motivated initial condition, the analysis performed in \cite{AfshordiManual} does not compare the effects of different orbital motions.

In this very exciting time for GW astronomy, some groups have started to search for echoes on the LIGO-Virgo available data set. Most of these searches were based on a matched-filtering method \cite{AfshordiDetection,AfshordiDetection1,AfshordiDetection3,WesterweckDetection,tsang2019morphologyindependent,Ricotemplate,UchikataDetection}, which requires \textit{a priori} knowledge of an accurate template, whereas some used morphology-independent methods  \cite{AfshordiDetection2,Holdomnewwindows,HoldomDectction,Salemi}. To this day, there are divergent claims about the existence of evidence for echoes in the data. For example, evidence for an echo detection with a significance of $4.2\sigma$ was reported in \cite{AfshordiDetection2}. On the other hand, the authors of \cite{WesterweckDetection} and \cite{UchikataDetection} claim to find no evidence supporting such detection. For a more detailed review of the search for echoes and the connection to quantum gravity, we direct the reader to \cite{AfshordiStatus, AfshordiQuantumReview, CardosoReviewECOS}. 

In \cite{Salemi}, several tentative post-merger signals in LIGO-Virgo data were reported with varying confidence levels. In \cite{AfshordiStatus}, some of these signals were considered as possible echo observations, proposing a correlation between their p-values and the mass ratio of the original binary: post merger signals from binaries with smaller mass ratios were considered to be more statistically significant (smaller p-value). These findings raise the question of whether there is a physical correlation between the mass ratio of the original binary and the echo amplitudes. It was discussed in \cite{AfshordiStatus} that although these data may suggest that binaries with more similar masses lead to smaller echo amplitudes, this conclusion may be a consequence of the coherent wave burst search pipeline used by Salemi et al. \cite{Salemi}.

In this work, we present estimates for the \emph{detectability} of the first echo by putting together two ingredients needed to model the echo excitation. First, we use a physically motivated expression for the ECO reflectivity \cite{Afshordi2}, within a range that brackets our uncertainty \cite{AfshordiSeismology} (see Section \ref{Setup}).
Second, in Section \ref{Orbits} we include a  more accurate prescription of the inspiral orbital motion incorporating the back reaction due to GW emission, based on \cite{Thorneorbits}. Our results for the echo properties and detectability are presented in Sections \ref{mudependence} and \ref{sec:detect} respectively, and we state our final conclusions in Section \ref{conclusions}. Throughout this work, we use units such that $c=G=\hbar=k_{B}=M=1$.

\section{Setup}\label{Setup}

We model ECOs by approximating their surrounding spacetime with the Kerr metric. This approximation is justified, for example, by the results found in \cite{PosadaMomentumInertia}, according to which the mass quadrupole moment of a slowly uniform density rotating star and its normalized moment of inertia approach the values for the Kerr metric as its compactness increases. Therefore the spacetime surrounding an ECO of spin $a$ and mass $M$ is assumed to be well-described by the following line element:

\begin{align}
&ds^{2} = -\dfrac{\Delta}{\Sigma} (dt-a \sin(\theta)d\phi)^{2} \nonumber +\\&+\dfrac{\sin^{2}(\theta)}{\Sigma} \left((r^{2}+a^{2})d\phi-adt\right)^{2}  + \dfrac{\Sigma}{\Delta}dr^{2} + \Sigma d\theta^{2},
\label{eq:Kerr}
\end{align}
where we use the definitions $\Delta\equiv r^{2} -2Mr +a^{2}$, $\Sigma\equiv r^{2}+a^{2}\cos^{2}(\theta)$ and the usual Boyer-Lindquist coordinate system. In our case of study, we consider that the spacetime is modified at the near-horizon region by the introduction of an effective reflective wall at $r^{0}=r_{+}(1+\epsilon)$ with $\epsilon \ll 1$, where $r_{+}\equiv M+\sqrt{M^{2}-a^{2}}$ is the expected position for the Kerr BH event horizon. It has been conjectured that $\epsilon$ is related to the Planck length $\ell_{Planck}$. This simplified model has been extensively used in other studies \cite{Mark:2017dnq,Micchi1,Holdomnewwindows,Holdomnewwindows,Paniquench,PaniAnalytical}. 

\subsection{Teukolsky formalism}

 Given this approximate description of an ECO spacetime, it is reasonable to assume that the Kerr perturbation equations also describe the perturbations around ECOs. This argument means that we are allowed to work with the Teukolsky equation \cite{Teukolsky1,Teukolsky2,Teukolsky3}, which for a given mode $(l,m)$ of frequency $\omega$ reads:  
\begin{align}
&\left( \dfrac{K^{2}-2\textit{i}s(r-M)K}{\Delta} + 4\textit{i}s\omega r - {}_{s}\lambda_{lmc} \right){}_{s}R_{lm\omega}(r) + \nonumber \\
&+ \Delta^{-s}\dfrac{d}{dr}\left( \Delta^{s+1}\dfrac{d {}_{s}R_{lm\omega}(r)}{dr} \right) = T_{l m \omega}(r),
\label{eqteukolsky}
\end{align}
where we use the following definitions:
\begin{align}
c\equiv a\omega, \quad K \equiv (r^{2}+a^{2})\omega -am, \nonumber
\end{align}
${}_{s}\lambda_{lmc}$ is the eigenvalue of the spin-weighted spheroidal harmonic equation and  ${}_{s}S_{lmc}$ is the corresponding eigenfunction \cite{Marcangular}. Here, $T_{l m \omega}$ is the energy-momentum tensor that acts as the source for the Teukolsky equation. All of our numerical results are for the mode $l=m=2$.

We restrict our study to the gravitational case ($s=-2$), in which the Teukolsky equation (\ref{eqteukolsky}) dictates the radial behaviour (${}_{-2}R_{lm\omega}$) of the Newman-Penrose scalar $\psi_{4}$ \cite{Newman1,Newman2}. For $r\rightarrow \infty$ it relates to the amplitudes of the plus $h_{+}$ and the cross $h_{\times}$ polarization modes of the GW as \cite{SasakiReview}:
\begin{equation}
     \psi_{4} = \dfrac{1}{2} (\ddot{h}_{+}-i \ddot{h}_{\times}), 
\end{equation}
and from now on we refer to the quantity
\begin{equation}
\label{eq:strain}
     h(t) = \dfrac{1}{2}(h_{+}-i h_{\times}), 
\end{equation}
as the \textit{strain} of the GW.

\begin{figure}
    \begin{center}
    \includegraphics[width=0.45\textwidth]{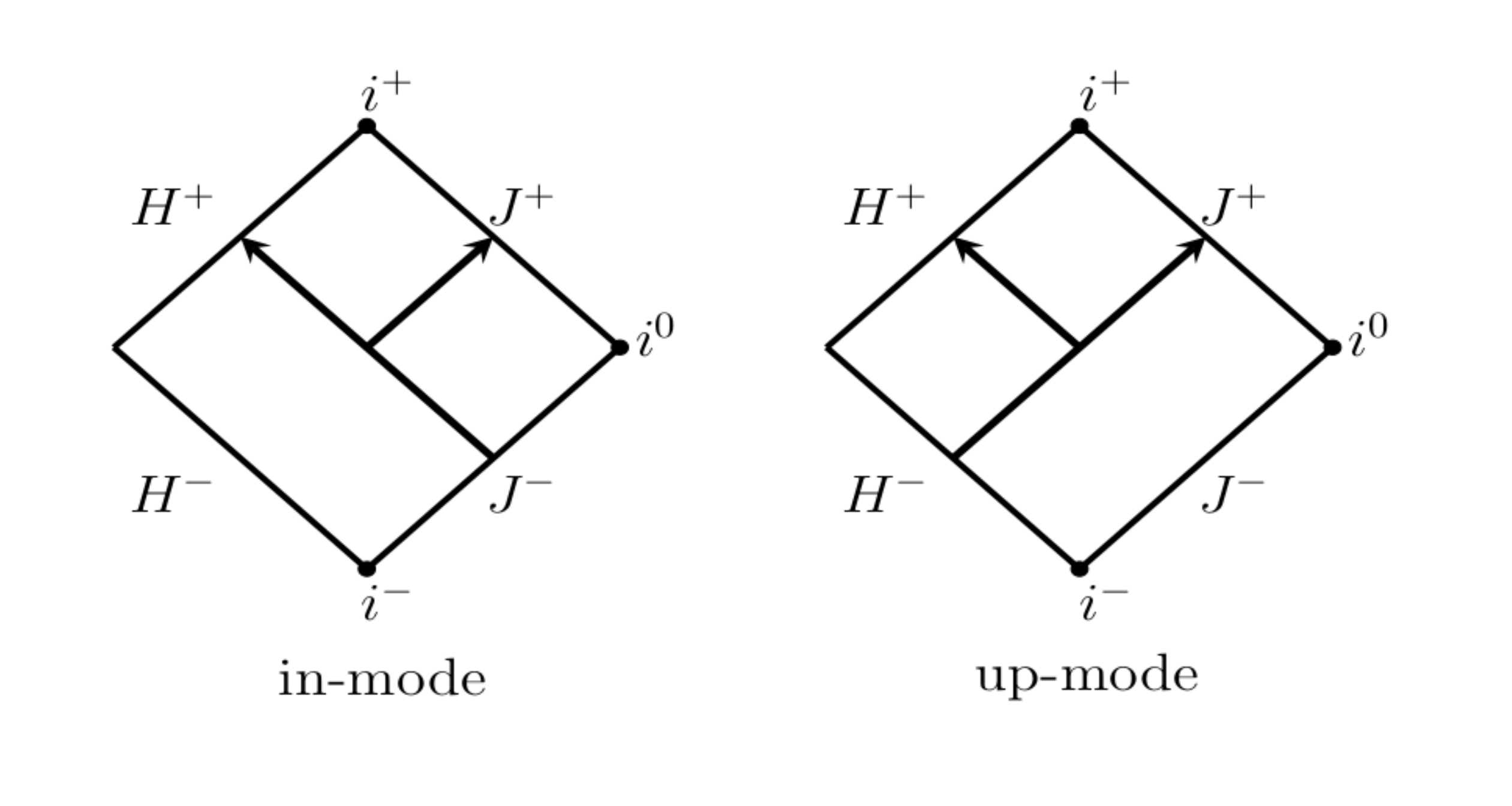}
  \end{center}
\caption{Representation of the boundary conditions of the \textit{in} and 
\textit{up-mode} solutions of the radial Teukolsky equation, given by 
eqs.~(\ref{eqin}) and (\ref{equp}), respectively, in a Penrose diagram.}
\label{fig:penrose}
\end{figure}

To solve the differential equation (\ref{eqteukolsky}), we use the Green's function technique. Therefore, two linearly independent homogeneous solutions are needed. The chosen solutions are the usual \emph{in-} and \emph{up-going} solutions (see Figure \ref{fig:penrose}). Their asymptotic behavior reads
\begin{equation}\label{eqin}
R_{lm\omega}^{\rm in}(r) \sim \left\lbrace \begin{array}{lr} 
B^{\rm trans}_{lm\omega}\Delta^{2}e^{-\textit{i}kr_{*}}, & \text{for} \quad r\rightarrow r_{+}, \\  \\
B^{\rm ref}_{lm\omega}r^3 e^{\textit{i}\omega r_{*}} + B^{\rm inc}_{lm\omega}\dfrac{e^{-\textit{i}\omega r_{*}}}{r}, &\text{for} \quad r\rightarrow \infty, 
\end{array} \right.
\end{equation}
where $B^{\rm trans/ref/inc}_{lm\omega}$ are the transmited/ reflected/ incident amplitudes of the in-going mode and
\begin{equation}\label{equp}
R_{lm\omega}^{\rm up}(r) \sim \left\lbrace \begin{array}{lr} 
C^{\rm ref}_{lm\omega}\Delta^{2} e^{-\textit{i} k r_{*}} + C^{\rm inc}_{lm\omega}e^{\textit{i} k r_{*}}, & \text{for} \quad r\rightarrow r_{+},\\ \\ 
C^{\rm trans}_{lm\omega}r^3 e^{\textit{i}\omega r_{*}}, & \text{for} \quad r\rightarrow \infty,
\end{array} \right.
\end{equation}
where $C^{\rm trans/ref/inc}_{lm\omega}$ are the transmitted/ reflected/ incident amplitudes of the up-going mode, $k \equiv \omega - m \Omega_{H}$ is the wave frequency at the horizon and $\Omega_{H}\equiv a/(2Mr_{+})$ is the angular velocity of the BH horizon. We also take the usual definition of the tortoise coordinate $r_{*}$, $ d r_{*}/dr = r^{2}+a^{2}/\Delta$, where we choose the integration constant such that:  

\begin{equation}
    r_{*}= r + \dfrac{ 2Mr_{+}}{r_{+}-r_{-}}\ln \left(\frac{r-r_{+}}{2M}\right) - \dfrac{ 2Mr_{-}}{r_{+}-r_{-}}\ln \left(\dfrac{r-r_{-}}{2M}\right).
\label{rstar}
\end{equation}

We follow the methodology described in \cite{SasakiReview} to construct the asymptotic amplitudes at the horizon and at infinity  of the inhomogeneous solution of the Teukolsky equation as: 

\begin{align}\label{Zinfsm2}
 & Z^{\rm BH,\infty}_{lm\omega} \equiv \frac{1}{2i\omega B^{\rm inc}_{lm\omega}} \int_{r_{+}}^{\infty}dr R^{\rm in}_{lm\omega}(r)\dfrac{T_{lm\omega}(r)}{\Delta(r)^{2}},
\end{align}
\begin{align}\label{ZHsm2}
 & Z^{\rm BH,H}_{lm\omega} \equiv\frac{B^{\rm trans}_{lm\omega}}{2i\omega B^{\rm inc}_{lm\omega} C^{\rm trans}_{lm\omega}}\int_{r_{+}}^{\infty}dr R^{\rm up}_{lm\omega}(r)\dfrac{T_{lm\omega}(r)}{\Delta(r)^{2}}, 
\end{align}
 where $T_{lm\omega}$ is the mode decomposition of the stress energy-tensor. As the expressions for $T_{lm\omega}$ are standard in the literature we choose not to show them here for the sake of brevity. We direct the reader to \cite{SasakiReview} for a detailed presentation of the equations. 
The inhomogenous solution of the Teukolsky equation for a BH is given by:
 \begin{equation}\label{Teukinhomo}
R_{lm\omega}^{\rm inhomo}(r) \sim \left\lbrace \begin{array}{lr} 
Z^{\rm BH,H}_{lm\omega}\Delta^{2}e^{-\textit{i}kr_{*}}, & \text{for} \quad r\rightarrow r_{+},\\ \nonumber \\
Z^{\rm BH,\infty}_{lm\omega} r^{3}e^{\textit{i}\omega r_{*}}, &\text{for} \quad r\rightarrow \infty. 
\end{array} \right.
\end{equation}

We find that the integral appearing in equation (\ref{ZHsm2}) has poor numerical convergence in the near horizon regime (but see also \cite{sago2020gravitational}). For this reason it is useful to employ the inhomogeneous solution of the Teukolsky equation at infinity (\ref{Zinfsm2}) as  an intermediate step to obtain the inhomogeneous solution of the Sasaki-Nakamura (SN) equation at the horizon (see equation (4.5) in \cite{SasakiOriginal}). Then we can obtain the horizon wave via an approximative scheme.

\subsection{Transforming to the Sasaki-Nakamura formalism}

For the SN equation \cite{SasakiOriginal}, the two homogeneous solutions with the same boundary condition as in (\ref{eqin}) and (\ref{equp}) are given by: 

\begin{equation}\label{eqinSN}
X_{lm\omega}^{\rm in}(r) \sim \left\lbrace \begin{array}{l} 
{}_{SN}B^{\rm trans}_{lm\omega}e^{-\textit{i}kr_{*}},  
\quad\quad\quad\quad\text{for} \quad r\rightarrow r_{+},\\ \\
{}_{SN}B^{\rm ref}_{lm\omega}e^{\textit{i}\omega r_{*}}\\ \quad\quad\quad\quad + {}_{SN}B^{\rm inc}_{lm\omega}e^{-\textit{i}\omega r_{*}}, \text{for} \quad r\rightarrow \infty,
\end{array} \right.
\end{equation}
\begin{equation}\label{equpSN}
X_{lm\omega}^{\rm up}(r) \sim \left\lbrace \begin{array}{l} 
{}_{SN}C^{\rm ref}_{lm\omega} e^{-\textit{i} k r_{*}} \\ \quad\quad\quad\quad+ {}_{SN}C^{\rm inc}_{lm\omega}e^{\textit{i} k r_{*}},
\text{for} \quad r\rightarrow r_{+}, \\ \\ 
{}_{SN}C^{\rm trans}_{lm\omega}e^{\textit{i}\omega r_{*}},\quad\quad\quad\quad\text{for} \quad r\rightarrow \infty.
\end{array} \right.
\end{equation}
For later convenience, we choose to normalize these solutions in the following way: 
\begin{equation}
    \bar{X}_{lm\omega}^{\rm up}(r) = \dfrac{X_{lm\omega}^{\rm up}(r)}{{}_{SN}C^{\rm trans}_{lm\omega}},
\end{equation}
\begin{equation}
    \bar{X}_{lm\omega}^{\rm in}(r) = \dfrac{X_{lm\omega}^{\rm in}(r)}{{}_{SN}B^{\rm trans}_{lm\omega}}.
\end{equation}

For the inhomogeneous solution of the SN with a reflective boundary condition near the horizon, one can make a different choice of independent homogeneous solutions. We choose to use $\bar{X}_{lm\omega}^{\rm up}$ and construct $\bar{X}_{lm\omega}^{\rm ECO}$ as

\begin{equation}\label{Xeco}
    \bar{X}_{lm\omega}^{\rm ECO}(r) = \bar{X}_{lm\omega}^{\rm in}(r)+ K_{lm\omega} \bar{X}_{lm\omega}^{\rm up}(r).
\end{equation}
We refer to $K_{lm\omega}$ as the {\it transfer function}. For the construction of $K_{lm\omega}$, we require that the in-going and the out-going fluxes of $\bar{X}_{lm\omega}^{\rm ECO}$ are proportional to each other. From this flux consideration, we impose that:

\begin{align} \label{requirement}
    \dfrac{1}{|b_{0}|} K_{lm\omega} \dfrac{{}_{SN}C^{\rm inc}}{{}_{SN}C^{\rm trans}} = & \nonumber \\= {\cal R} e^{-2ikr_{*}^{0}}&\dfrac{1}{|C|}\left(1+ K_{lm\omega} \dfrac{{}_{SN}C^{\rm ref}}{{}_{SN}C^{\rm trans}_{lm}}\right).
\end{align}
The definitions of $b_{0}$ and $C$ and more details about the energy fluxes can be found in Appendix \ref{appendixfluxes}, and ${\cal R}$ is defined as the reflectivity of the ECO surface. The above condition allows us to determine\footnote{Given the same choice of reflectivity, our transfer function would have the same pole structure as the one described in \cite{Holdomnewwindows, HoldomWaves}. In this case we expect that both models present the same QNM spectrum for the ECO.} $K_{lm\omega}$ as: 
\begin{eqnarray}\label{KSN}
K_{lm\omega}&=&\dfrac{|b_{0}|}{|C|}\frac{{}_{SN}C^{\rm trans}_{lm\omega}}{{}_{SN}C^{\rm inc}_{lm\omega}} \frac{{\cal R} e^{-2i k r_{*}^{0}}}{1- \dfrac{|b_{0}|}{|C|}\dfrac{{}_{SN}C^{\rm ref}_{lm\omega}}{{}_{SN}C^{\rm inc}_{lm\omega}}{\cal R}e^{-2i k r_{*}^{0}}} \\
&=& -\dfrac{|b_{0}|}{|C|}\frac{c_{0}C^{\rm trans}_{lm\omega}}{4 \omega^{2}g C^{\rm inc}_{lm\omega}} \frac{{\cal R} e^{-2i k r_{*}^{0}}}{1- \dfrac{|b_{0}|}{|C|}\dfrac{dC^{\rm ref}_{lm\omega}}{gC^{\rm inc}_{lm\omega}}{\cal R}e^{-2i k r_{*}^{0}}}.\nonumber
\label{Komegabarsm2}
\end{eqnarray} 
In the previous equation, we use the quantities $c_{0},d,g$, which are defined in Appendix \ref{appendixA}. In our numerical analysis, all physical quantities are obtained through the MST method \cite{MSToriginal,SasakiReview,Micchi1}, i.e. in the Teukolsky formalism. They are later transformed to the SN formalism by means of standard relations \cite{SasakiOriginal,SasakiReview,Holdomnewwindows}, also summarized in Appendix \ref{appendixA}.  
Once the transfer function is found, we can construct the ECO Green's function as:

\begin{align}
G^{\rm ECO}(r|r') &= G^{\rm BH}(r|r') + K_{lm\omega} \dfrac{ \bar{X}_{lm\omega}^{\rm up}(r)\bar{X}_{lm\omega}^{\rm up}(r')}{W_{lm\omega}},
\label{GFECO}
\end{align}
where 
\begin{align}\label{GF}
G^{\rm BH}_{lm\omega}(r|r') =& \ \dfrac{\bar{X}_{lm\omega}^{\rm up}(r)\bar{X}_{lm\omega}^{\rm in}(r')}{W_{lm\omega}}\Theta(r-r')\nonumber +\\&+ \dfrac{\bar{X}_{lm\omega}^{\rm up}(r')\bar{X}_{lm\omega}^{\rm in}(r)}{W_{lm\omega}}\Theta(r'-r),
\end{align}
 is the usual BH Green's function and $W_{lm\omega}$ is the Wronskian between $\bar{X}_{lm\omega}^{\rm up}$ and $\bar{X}_{lm\omega}^{\rm in}$.
 
 Integrating against the source term, one can easily show that, for the BH case, the perturbative response has as general form: 
\begin{equation}\label{SNinhomo}
X_{lm\omega}^{BH}(r) \sim \left\lbrace \begin{array}{lr} 
X^{\rm BH,H}_{lm\omega} e^{-\textit{i} k r_{*}}, & \text{for} \quad r\rightarrow r_{+}, \\ \nonumber \\
X^{\rm BH,\infty}_{lm\omega}e^{\textit{i}\omega r_{*}}, & \text{for} \quad r\rightarrow \infty.
\end{array} \right.
\end{equation}

On the other hand, for the ECO case, the general solution  in the asymptotic regime $r\rightarrow \infty$ will be given by:
\begin{equation}
X_{lm\omega}^{ECO}(r) \sim 
(X^{\rm BH,\infty}_{lm\omega}(r) +K_{lm\omega}X^{\rm BH,H}_{lm\omega}(r)) e^{\textit{i}\omega r_{*}},
\label{ISN}
\end{equation}
meaning that the ECO's perturbative response can be modeled if one knows the BH's response both near and far from the horizon ($X^{\rm BH,H}_{lm\omega}$ and $X^{\rm BH,\infty}_{lm\omega}$ respectively). 

 In \cite{SasakiOriginal} it was shown that, although the general transformation of the inhomogeneous solution is not the same as for the homogeneous case, the relation between $X^{\rm BH,\infty}$ and $Z^{\rm BH,\infty}$ is the same as in (\ref{inrealtionsb}) (see equations (2.19) and (4.5) in \cite{SasakiOriginal}). This means that for the BH case the perturbative response at infinity, in the SN formalism, will be given by:

\begin{align}\label{XinfSN}
 X^{\rm BH,\infty}_{lm\omega} = -\dfrac{c_{0}}{4\omega^{2}} Z^{\rm BH,\infty}_{lm\omega},
\end{align}
where $Z^{\rm BH,\infty}_{lm\omega}$ is given in the Teukolsky formalism by equation (\ref{Zinfsm2}). Even if we were able to find a transformation between $X^{\rm BH,H}_{lm\omega}$ and $Z^{\rm BH,H}_{lm\omega}$ similar to the one in (\ref{XinfSN}), the integral appearing in (\ref{ZHsm2}) is (numerically) poorly convergent near the horizon. For this reason, we choose to use an approximation for calculating $X^{\rm BH,H}_{lm\omega}$ in terms of $X^{\rm BH,\infty}_{lm\omega}$ such as described in Appendix B of \cite{PaniAnalytical} (see eq. (B7) in \cite{PaniAnalytical}) which for our case can be translated as: 
\begin{align}\label{XhSN}
 X^{\rm BH,H}_{lm\omega} &\approx \dfrac{{}_{SN}B^{\rm trans}_{lm\omega}}{{}_{SN}B^{\rm ref}_{lm\omega}} X^{\rm BH,\infty}_{lm\omega} \nonumber \\  &=  \left(\dfrac{4\omega^{2}dB^{\rm trans}_{lm\omega}}{c_{0}B^{\rm ref}_{lm\omega}}\right)\dfrac{c_{0}}{4\omega^{2}} Z^{\rm BH,\infty}_{lm\omega}.
\end{align}
Two points should be made about the use of this approximation. First, (\ref{XhSN}) is a good approximation for frequencies close to the BH fundamental QNM frequency. We note that, for the particular model investigated here, most of the frequencies expected to be excited are close to the QNM frequency due to our choice of reflectivity ${\cal R}$ (see Section \ref{secref}). Second, in \cite{PaniAnalytical} this relation was derived for the Chandrasekhar-Detweiler equation. However, similar  considerations apply to our case, because both equations have homogeneous solutions that behave as plane waves at $r \rightarrow r_{+}$ and $r \rightarrow \infty$. Therefore, it is straightforward to show that the coefficient relating the waves at the horizon and at infinity is the ratio between the asymptotic amplitudes. This is the same reason why it is not possible to find a similar approximation in the Teukolsky formalism (see Equations (\ref{eqin}) and (\ref{equp})).

Therefore the final ECO response, at $r\rightarrow \infty$, will be given by:
\begin{align}\label{final}
& X_{lm\omega}^{ECO}(r) \sim -\dfrac{c_{0}}{4\omega^{2}} \times \nonumber\\ 
& \times \left[1-K_{lm\omega}\left(\dfrac{4\omega^{2}dB^{\rm trans}_{lm\omega}}{c_{0}B^{\rm ref}_{lm\omega}}\right)\right]Z^{\rm BH,\infty}_{lm\omega}
 e^{\textit{i}\omega r_{*}}.
\end{align}
In order to obtain the strain $h(t)$ (\ref{eq:strain}) from equation (\ref{final}) we need only drop the overall factor $-c_{0}/4$ and perform an inverse Fourier transform \cite{SasakiOriginal}, taking into account the inclination angle of the binary (assumed here to be face-on) and the distance to the source as needed. 

\subsection{Surface Reflectivity}\label{secref}

While most of the works in the literature have been restricted to the constant reflectivity case (e.g., \cite{PaniAnalytical,Micchi1,Holdomnewwindows}), we focus on the case of Boltzmann reflectivity, described for the first time in this context in \cite{Afshordi2}. This reflectivity is given by the following expression: 
\begin{equation}\label{RBoltz}
    {\cal R}_{B}=e^{-\dfrac{|k|}{2 \alpha T_{\rm H}}},
\end{equation}
where
\begin{equation} \label{temp}
 T_{\rm H}=\dfrac{1}{4M\pi} \dfrac{\sqrt{1-(a/M)^{2}}}{1+\sqrt{1-(a/M)^{2}}},
\end{equation}
is the BH Hawking temperature. One of the reasons for choosing this prescription for the reflectivity is that it provides a natural cut-off for frequencies that deviate from the BH fundamental QNM, providing a relatively safe regime for using the approximation (\ref{XhSN}). 
Arguments based on the assumptions of CP-symmetry, fluctuation-dissipation theorem, or detailed balance in Rindler geometry each independently lead to the same law of Boltzmann energy reflectivity (\ref{RBoltz}) with $\alpha=1$ \cite{Afshordi2}. However, since near-horizon Rindler geometry may be modified by quantum effects, it was later suggested that a quantum BH may have a temperature higher than the classical BH by an overall factor of $\alpha \lesssim 2$ \cite{AfshordiSeismology} (to avoid the ergoregion instability). For this reason, we use two sample values of $\alpha=1$ (classical Boltzmann reflectivity) and $\alpha =2$ (the maximum value without ergoregion instability). The latter value should also maximize the echo reflectivity and echo amplitudes. 

\begin{figure*}[ht!]
	   \includegraphics[width=0.4\textwidth]{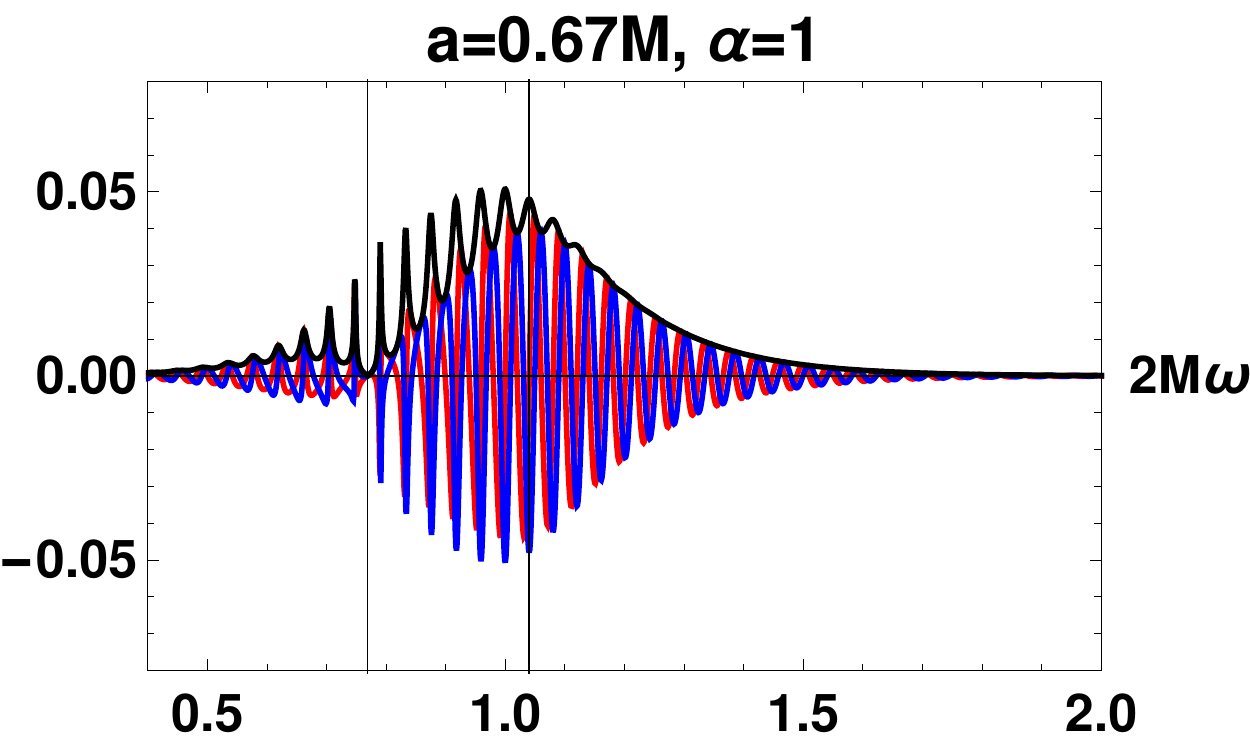}
	   \includegraphics[width=0.51\textwidth]{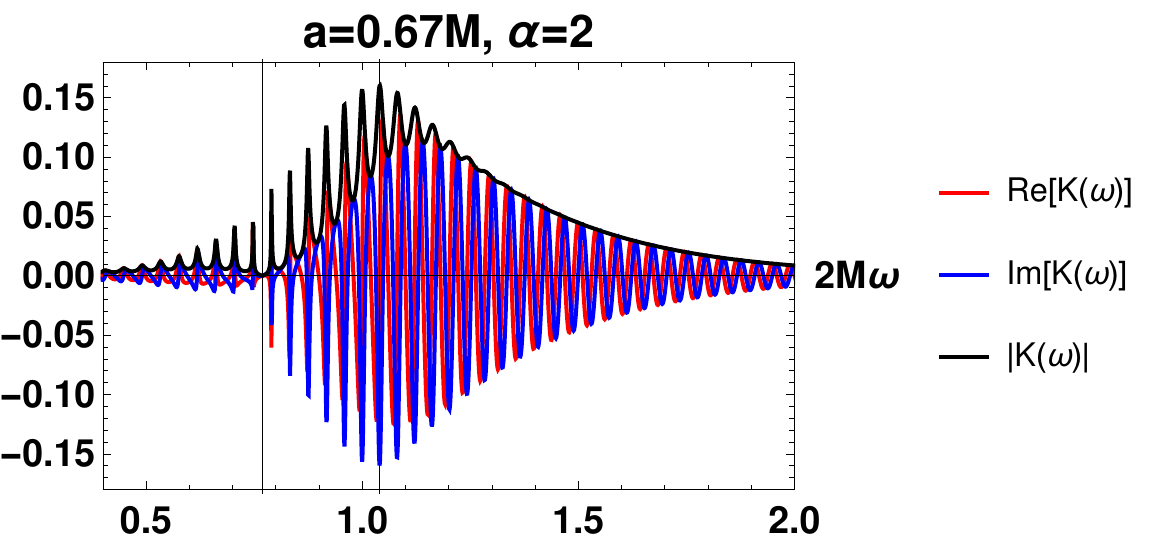}
	   \includegraphics[width=0.4\textwidth]{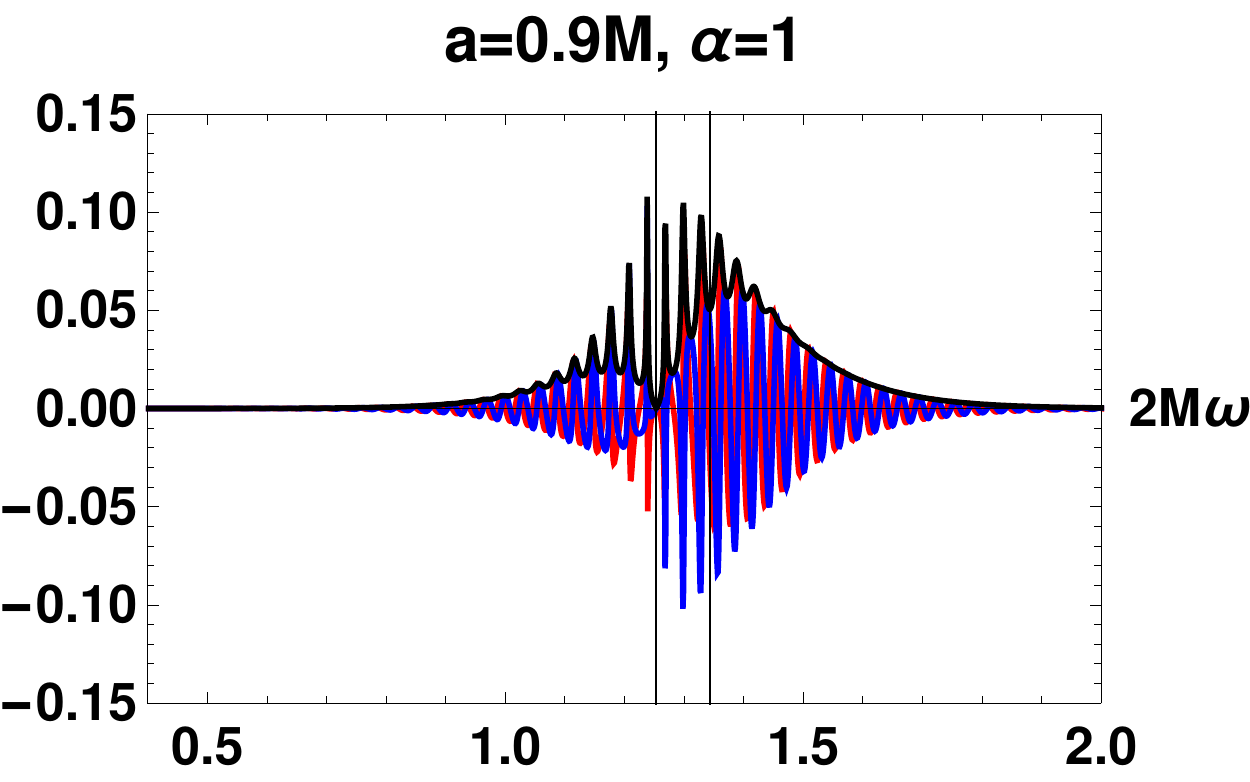}
	   \includegraphics[width=0.49\textwidth]{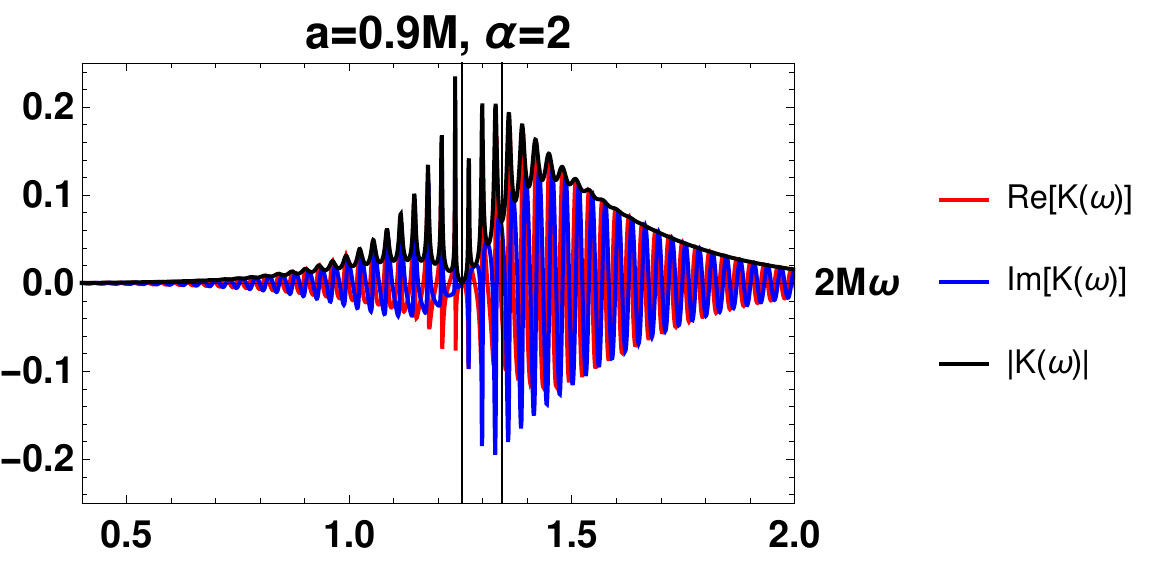}
\caption{ Transfer functions for the $l=m=2$ mode. In the top row the position of the reflective wall is $r_{*}^{0}=-150M$ and $a=0.67M$, while for the bottom row is $r_{*}^{0}=-210.5M$ and $a=0.9M$ (see Section \ref{spindependence} for the discussion of this choice of parameters) . The different columns represent different choices of $\alpha$. The vertical lines mark the position of the superradiant bound frequency and the fundamental BH QNM (from left to right). The transfer function goes to zero at $\omega_{SR}$ and the  superradiant frequencies ($\omega<\omega_{SR}$) are in general disfavored in comparison with non-superradiant ones ($\omega>\omega_{SR}$).}
\label{Ktransferfunctions}
\end{figure*}

Using this choice of reflectivity, we construct the transfer function and the results can be found in Figure \ref{Ktransferfunctions}.  The most striking feature of Figure \ref{Ktransferfunctions} is that the absolute value of $K_{lm\omega}$ goes to 0 at the superradiant bound frequency, $\omega_{SR} \equiv m \Omega_H$. This is due the fact that the $\omega_{SR}$ is a totally reflected frequency.\footnote{By definition $\omega_{SR}$ is the frequency at which the reflectivity of the BH angular momentum barrier is equal to 1. Lower (higher) frequencies have reflectivity larger (smaller) than 1 due to the superradiance phenomenon \cite{Brito:2015oca}.}  We can also see that the superradiant frequencies ($\omega<\omega_{SR}$) are (in general) harder to excite when compared with $\omega>\omega_{SR}$.

For the rapidly spinning ECO (shown in the bottom panels of Figure \ref{Ktransferfunctions}), the value of $K_{lm\omega}$ decays faster as $|\omega-\omega_{SR}|\rightarrow\infty$, but it also has higher values. These two competing features are in agreement with the results in \cite{Afshordi2}, where it was shown that the amplitude of echoes does not increase monotonically for increasing ECO spin. This behaviour is caused by the competing effects of the enhancement due to superradiance and the decrease due to $T_{\rm H}\rightarrow0$ as $a\rightarrow M$.

\section{Trajectories}\label{Orbits}

In order to evaluate the detectability of the echoes from ECOs in a reasonably realistic model, in addition to the reflectivity (\ref{RBoltz}) we need an approximate description of the binary inspiral, which provides the \emph{initial conditions} for the excitation of the echoes. 
To do so and take into account the effect from different binary mass ratios, we slightly modify the prescription for extreme-mass-ratio orbits described in \cite{Thorneorbits}.
The prescription for the orbital motion consists of three different stages: 1) an adiabatic inspiral; 2) a transition phase; and 3) a geodesic plunge when the innermost stable circular orbit (ISCO) is crossed. In this section, we give a brief overview of the method used to obtain these orbits. 

All quantities marked with an upper tilde are scaled by the central object's mass, i.e. $\tilde{a}\equiv a/M$ and $\tilde{\omega}\equiv M\omega$.  It is important to note that, as seen in \cite{Micchi1}, we expect differences between corotating and counterrotating orbits. In \cite{Micchi1}, these differences resulted from the asymmetry of the transfer function with respect to the $\omega=0$ axis. This asymmetry also exists here, but negative frequencies are strongly suppressed due to our choice of Boltzmann reflectivity (eq.(\ref{RBoltz})). Therefore, we expect the echoes to be orders of magnitude smaller for counterrotating orbits. For this reason we choose to focus only on the corotating orbits.

\subsection{Inspiral Phase}

For the inspiral phase of the orbital motion, we use an adiabatic evolution between circular orbits at the equatorial plane. In this approximation, a smaller body of mass $\mu = q M$ ($0<q<1$) at a distance $\tilde{r}$ from the central body of mass $M$ has orbital energy $E$ given by: 
\begin{equation}
    E = q M \dfrac{1-2/\tilde{r}+\tilde{a}/\tilde{r}^{3/2}}{\sqrt{1-3/\tilde{r}+2 \tilde{a}/\tilde{r}^{3/2}}}.
\end{equation}
Due to the radiation reaction, the particle will lose orbital energy in a rate equal to the energy emitted in GW:

\begin{equation}\label{energyloss}
    \dot{E}_{GW}=-\dot{E}=\dfrac{32}{5} q^{2}\tilde{\Omega}^{10/3}\dot{\mathcal{E}},
\end{equation}
where $\tilde{\Omega}$ is the orbital velocity of the motion and $\dot{\mathcal{E}}$ is a parameter motivated by corrections to the Newtonian quadrupole formula, which accounts for the energy flux due to the orbital motion and depends only on the central body's mass and spin. The values of $\dot{\mathcal{E}}$ used here were obtained by interpolating the entries found on Table I of \cite{Thorneorbits}. 

Furthermore, the angular velocity of the orbital motion is approximated by the circular orbit equation \cite{Hughes,Thorneorbits}: 

\begin{equation}\label{angularevolution}
    \tilde{\Omega}\equiv M\Omega  =\dfrac{d\phi}{d\tilde{t}} =\dfrac{1}{\tilde{r}^{3/2}+\tilde{a}}.
\end{equation}
With the orbital energy loss (\ref{energyloss}), the evolution of the radial coordinate satisfies  the differential equation:

\begin{equation}
    \dfrac{dr}{dt} = \dfrac{-\dot{E}_{GW}}{dE/dr},
\end{equation}
and the evolution of the time coordinate follows the circular geodesic approximation:

\begin{equation}\label{timeevolution}
\dfrac{d\tilde{t}}{d\tilde{\tau}} = \dfrac{1 + \tilde{a}/r^{3/2} }{\sqrt{1+ 3/\tilde{r} + 2\tilde{a}/r^{3/2}}}.
\end{equation}

\subsection{Transition regime}

The adiabatic evolution between circular geodesics described in the previous Section is no longer valid near the ISCO. 
From $r= r_{\text{\tiny{ISCO}}}+ q^{2/5}R_{0}$, where $R_{0}\equiv(\beta \kappa)^{2/5} \gamma^{-3/5}$, we describe the radial evolution with the inclusion of nondissipative self-force effects as in \cite{Thorneorbits}:

\begin{equation}
    \dfrac{d^{2}\tilde{r}}{d\tilde{\tau}^{2}} = - \gamma (\tilde{r}-\tilde{r}_{\text{\tiny{ISCO}}})^{2} - \beta q \kappa \tilde{\tau} ,
\end{equation}
where we use the definitions:
\begin{align}
\gamma=\dfrac{3}{\tilde{r}_{\text{\tiny{ISCO}}}^{6}}( \tilde{r}^{2} + 2(\tilde{a}^{2}(&\tilde{E}^{2}-1) -\tilde{L}^{2} ) \tilde{r} \nonumber \\   &+10(\tilde{L}-\tilde{a}\tilde{E})^{2})_{\text{\tiny{ISCO}}}, 
\end{align}

\begin{align}
\beta= \dfrac{2}{\tilde{r}_{\text{\tiny{ISCO}}}^{4}}( (&\tilde{L}-\tilde{a}^{2}\tilde{E}\tilde{\Omega} )\tilde{r}\nonumber \\ &- 3 (\tilde{L}-\tilde{a}\tilde{E})(1 -\tilde{a}\tilde{\Omega}))_{\text{\tiny{ISCO}}},
\end{align}
\begin{equation}
\kappa= \dfrac{32}{5}\tilde{\Omega}_{\text{\tiny{ISCO}}}^{7/3}\dfrac{1+\tilde{a}/\tilde{r}^{3/2}}{\sqrt{1-3/\tilde{r}_{\text{\tiny{ISCO}}}+ 2\tilde{a}/\tilde{r}_{\text{\tiny{ISCO}}}^{3/2}}} \dot{\mathcal{E}},
\end{equation}
\begin{align}
\tilde{E}= \dfrac{E}{q M}= \dfrac{1- 2/\tilde{r}+a/\tilde{r}^{3/2}}{\sqrt{1-3/\tilde{r}+2\tilde{a}/\tilde{r}^{3/2}}},
\end{align}
and
\begin{align}
\tilde{L}= \dfrac{L}{q M^{2}}= \dfrac{2}{\sqrt{3 \tilde{r}}}(3\sqrt{r}-2\tilde{a}).
\end{align}
However, unlike \cite{Thorneorbits}, in our prescription for this transition period we keep the angular and time evolution of the orbit as given by equations (\ref{angularevolution}) and (\ref{timeevolution}). This avoids nonphysical oscillations in the waveform found in the original Ori-Thorne prescription due to discontinuous velocities \cite{Surrogate}.

\subsection{Geodesics: Plunge Phase}

After crossing the ISCO, we assume that the energy radiated away due to the particle motion has negligible effect on the evolution of the orbit.  Therefore we can approximate the plunge by a geodesic motion starting at $r=r_{\text{\tiny{ISCO}}}$. The energy $E$ and angular momentum $L$ are fixed by requiring that the evolution of the particle's coordinates $(r, t, \phi)$ and their derivatives be continuous. In practice we only require the continuity of $(r, \phi)$ and their derivatives, as there are only two variables to fix $(E, L)$ and three equations to guarantee smoothness (the derivatives of the three coordinates $(r, t, \phi)$). We verified that imposing the smoothness condition to $r$ and $\phi$ leads to values of $E$ and $L$ which guarantee that the derivative of $t$ is discontinuous by less than one percent.

\begin{figure}[H]
 \centering	   \includegraphics[width=0.4\textwidth]{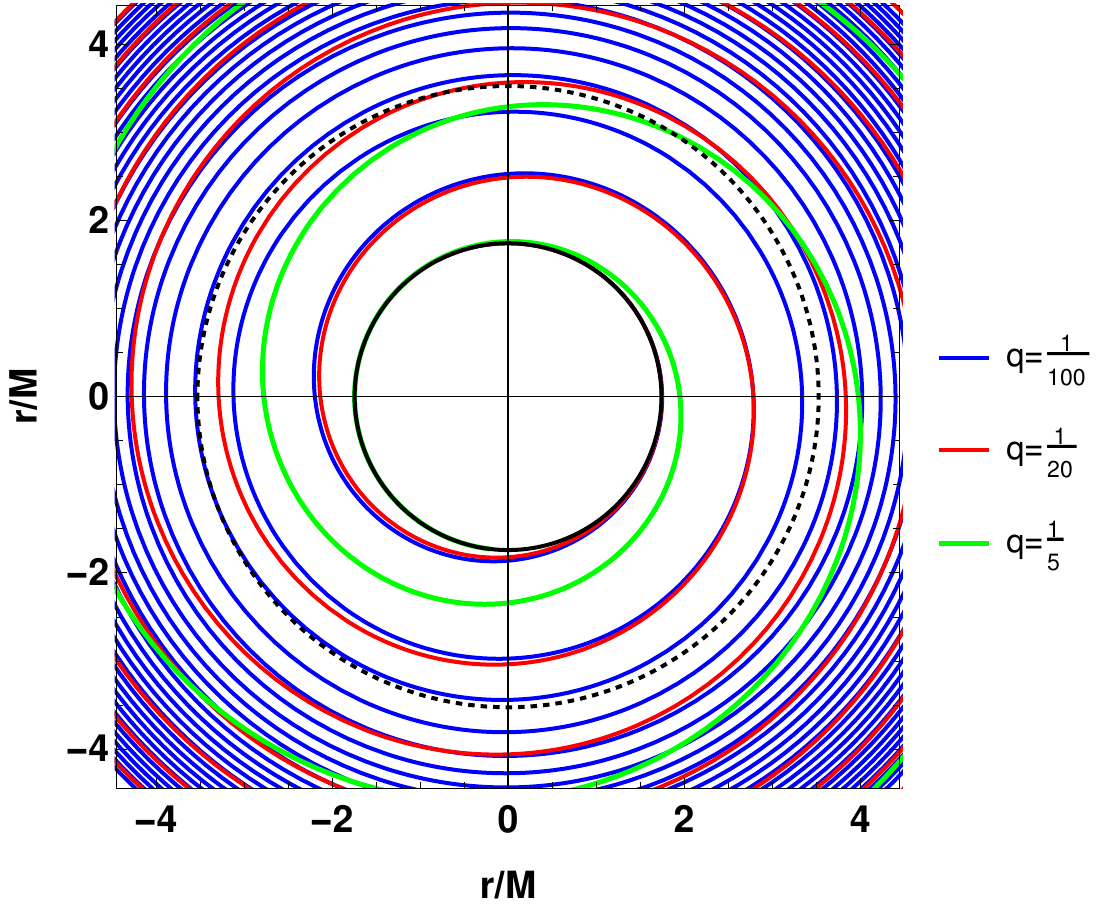}
\caption{Three examples of orbital motion. For all cases $a=0.67M$ and the colors represent different mass ratios $q=1/5$ (green), $1/20$ (red) and $1/100$ (blue). BH horizon and ISCO are shown in solid and dotted black lines, respectively. Orbits with smaller mass ratio evolve more slowly, even after crossing the ISCO.}
\label{orbitsa067M}
\end{figure}

In Figure \ref{orbitsa067M} we show three examples of orbital motions obtained with the prescription outlined in this Section. These orbits are, as required, smooth when crossing the ISCO. Even though geodesic motions are independent of the mass ratio, they are highly dependent on the initial conditions. In our case, the initial conditions for the free-fall are set by the first stages of the motion, which do depend on the mass ratio. As a result, we find that orbits with smaller mass ratio evolve more slowly, even after crossing the ISCO.

\section{Echo properties}\label{mudependence}

As discussed in Section \ref{intro}, there seems to be evidence supporting a correlation between the amplitude of the echoes and the mass ratio of the progenitor binary \cite{Salemi,AfshordiStatus}. To analyse this possible $q$-dependence, we fix the central object spin at $a=0.67M$, the position of the wall at $r_{*}^{0}=-150M$ and choose the Boltzmann reflectivity (\ref{RBoltz}) to be used in our transfer function (\ref{KSN}). We select five different values for the mass ratio, namely $q=1/100,1/50,1/20,1/10,1/5$. With this setup, we evolve the waveforms for the same time before the plunge ($\Delta t \sim 800M$).

We choose these values of $q$ for three main reasons. First, had we chosen smaller mass ratios the plunge would take longer to evolve with no particular gain of insight. Second, most of the current observations are for binaries of similar masses \cite{Ligo1, Ligo2, Ligo3, Ligo4, Ligo5, Ligo6, LigoCatalog}, with the notable exception of GW190412 \cite{Ligo8}. Third, it has been recently shown that, although not strictly in its regime of validity, the perturbative extreme mass ratio approximation performs surprisingly well for binaries of comparable masses \cite{Surrogate}.

In \cite{Surrogate}, it was shown that waveforms emitted by the coalescence of non spinning objects obtained within the linear regime for $ 3 \leqq q^{-1} \leqq 10 $ can be adjusted to match results from Numerical Relativity (NR). The proposed rescaling is given by: 
\begin{align} \label{rescale}
    h_{NR}(t;q) &\approx  \xi(\eta) h_{PT}(\xi(\eta) t;q),    
\end{align}  
\begin{align}
    \rm{with} \quad \eta &\equiv \dfrac{q}{(1+q)^2},
\end{align}  
where $h_{NR}(t;q)$ and $h_{PT}(t;q)$ are the waveforms as obtained by NR and perturbation theory, respectively, and $\xi(\eta)$ is a scaling function that depends only on the symmetric mass ratio $\eta$. In the limit of $\eta\rightarrow0$, the scaling factor should be $\xi = 1+q$ \cite{Surrogate}. However, away from the extreme mass ratio regime this factor can be approximated as a fourth order polynomial function of $\eta$, which is  monotonically decreasing in the interval $0<q<1$.
The results obtained for non-spinning binaries in \cite{Surrogate} should not hold quantitatively in our case. However, we expect a similar trend between our model and a (not yet available) NR simulation of echoes. As it is clear from equation (\ref{rescale}), the rescaling in amplitude is a global factor, as one expects from a change of an overall normalization. Therefore we do not expect any qualitative modification in the behaviour of the echo trends discussed in the following subsections.

 \subsection{Amplitude and peak time dependence on mass ratio} \label{amplitudesec}

\begin{figure*}[ht!]
\includegraphics[width=0.28\textwidth]{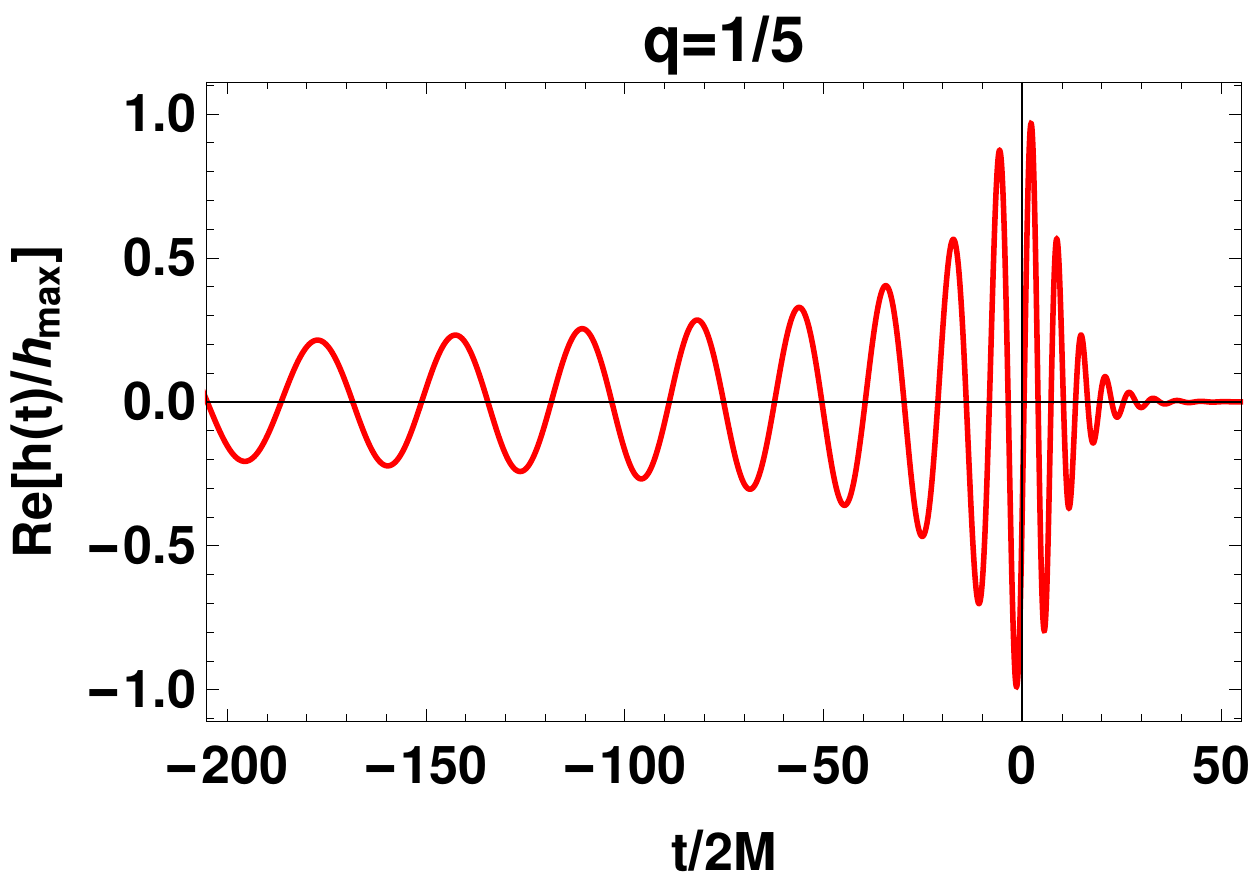}
\includegraphics[width=0.28\textwidth]{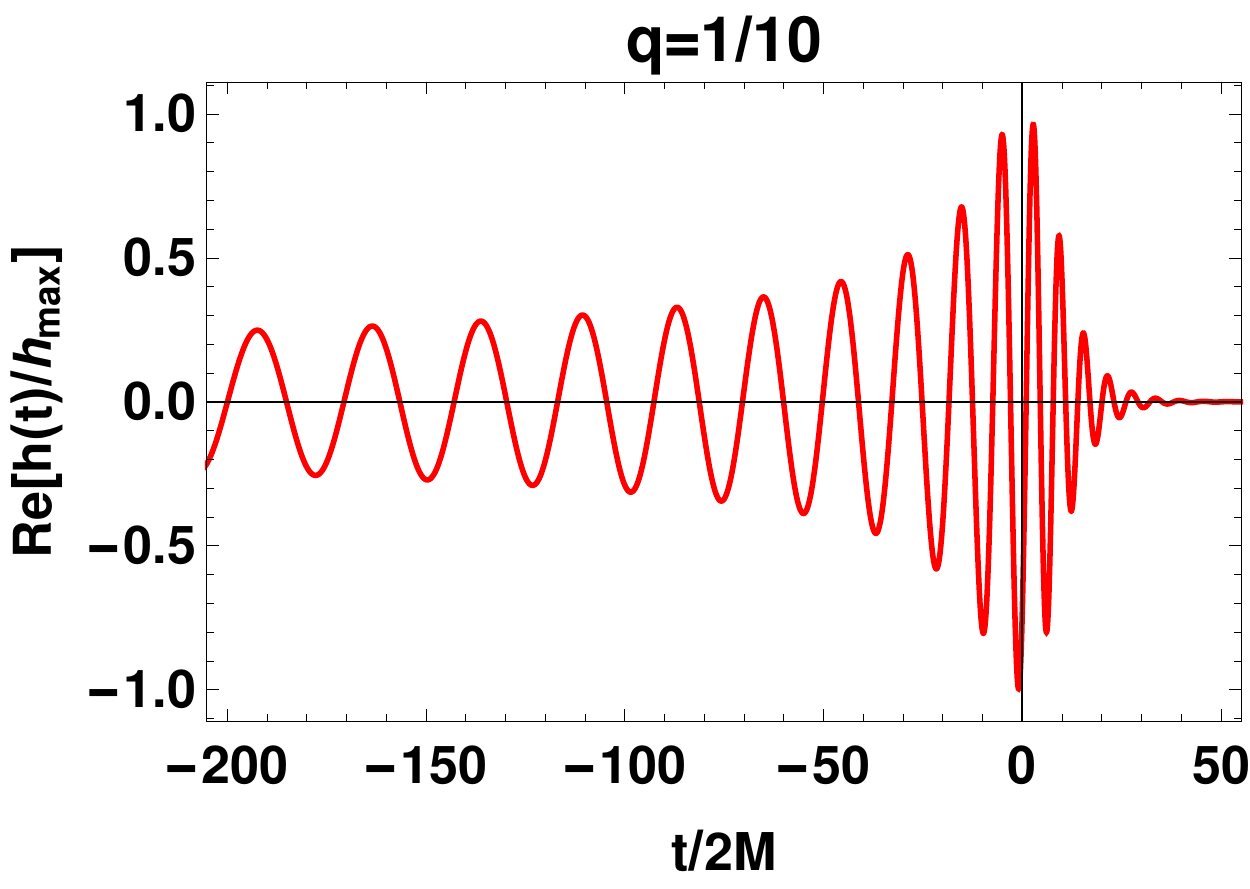}
\includegraphics[width=0.28\textwidth]{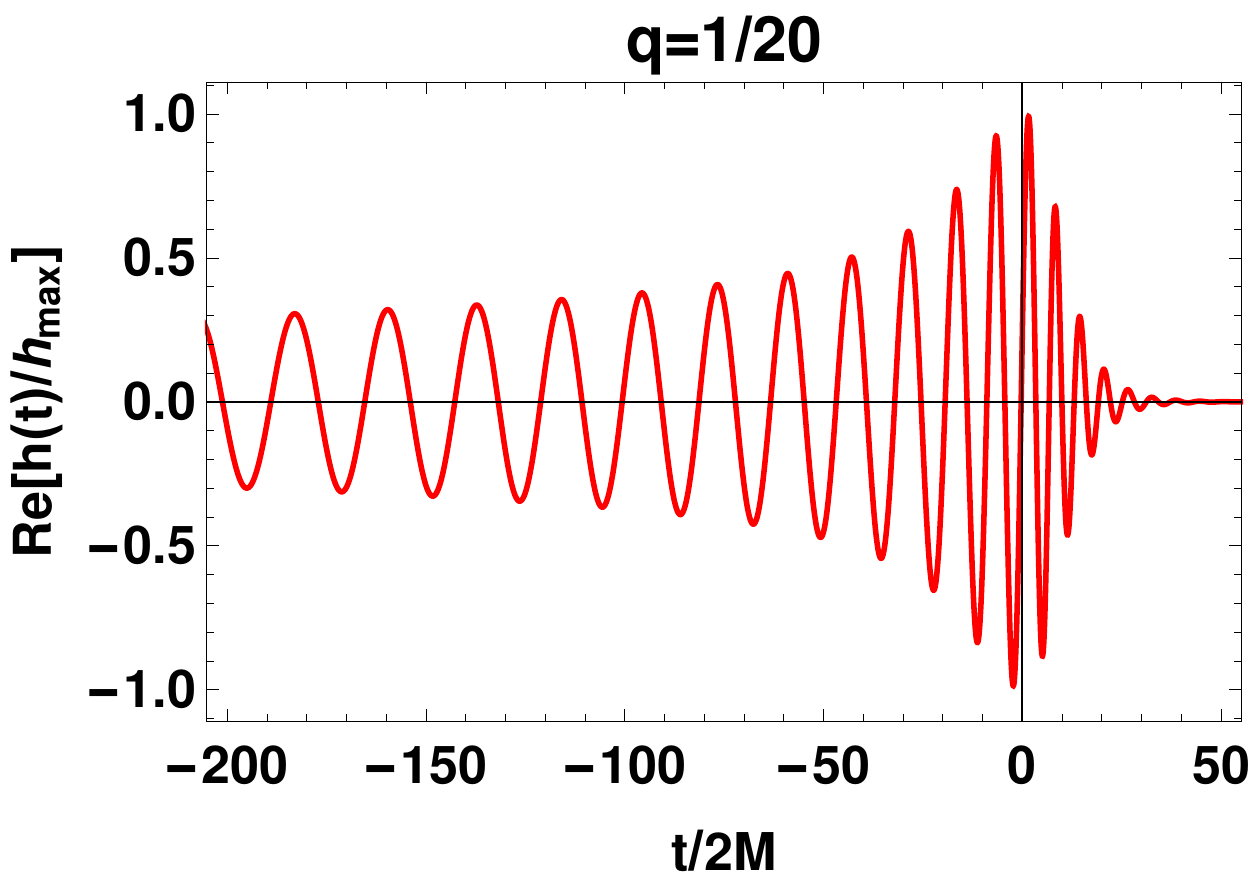}
\includegraphics[width=0.28\textwidth]{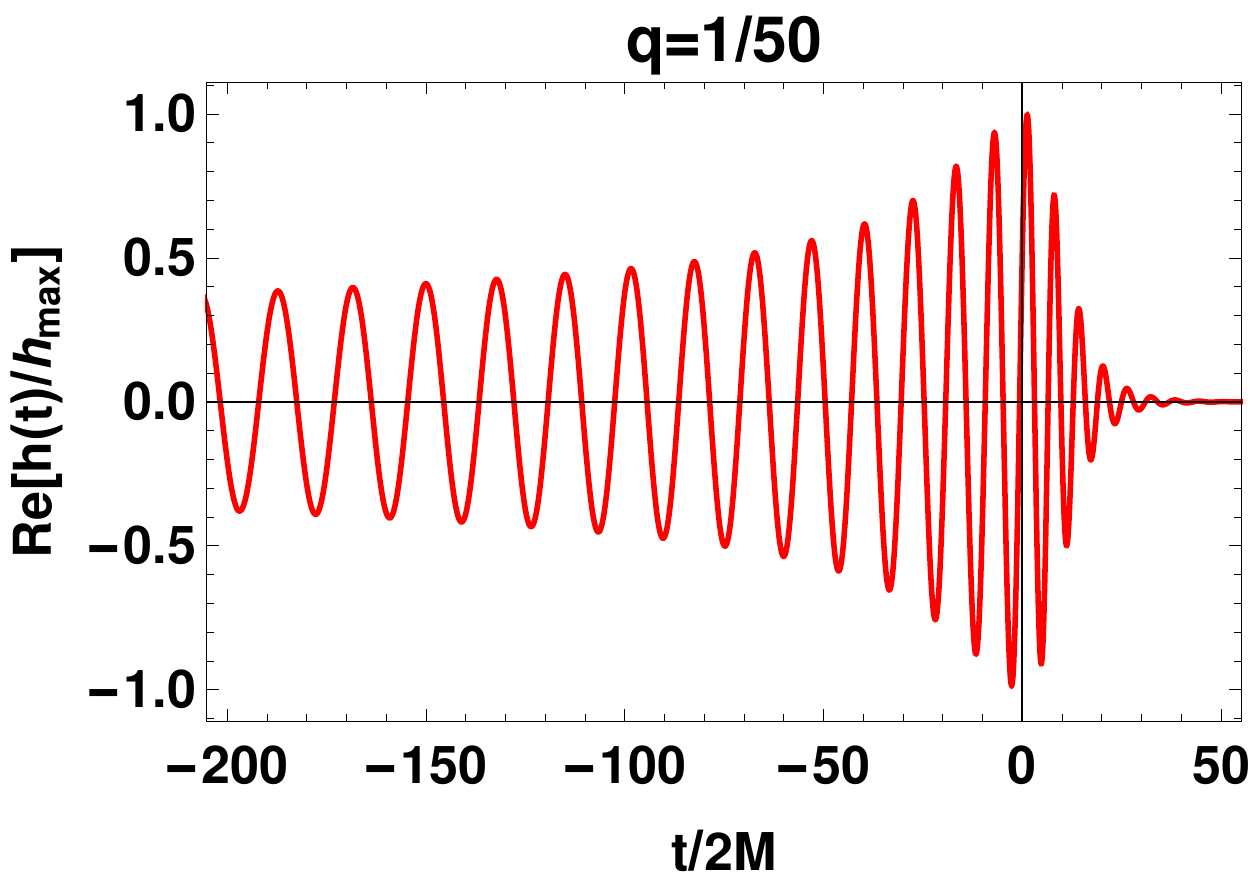}
\includegraphics[width=0.28\textwidth]{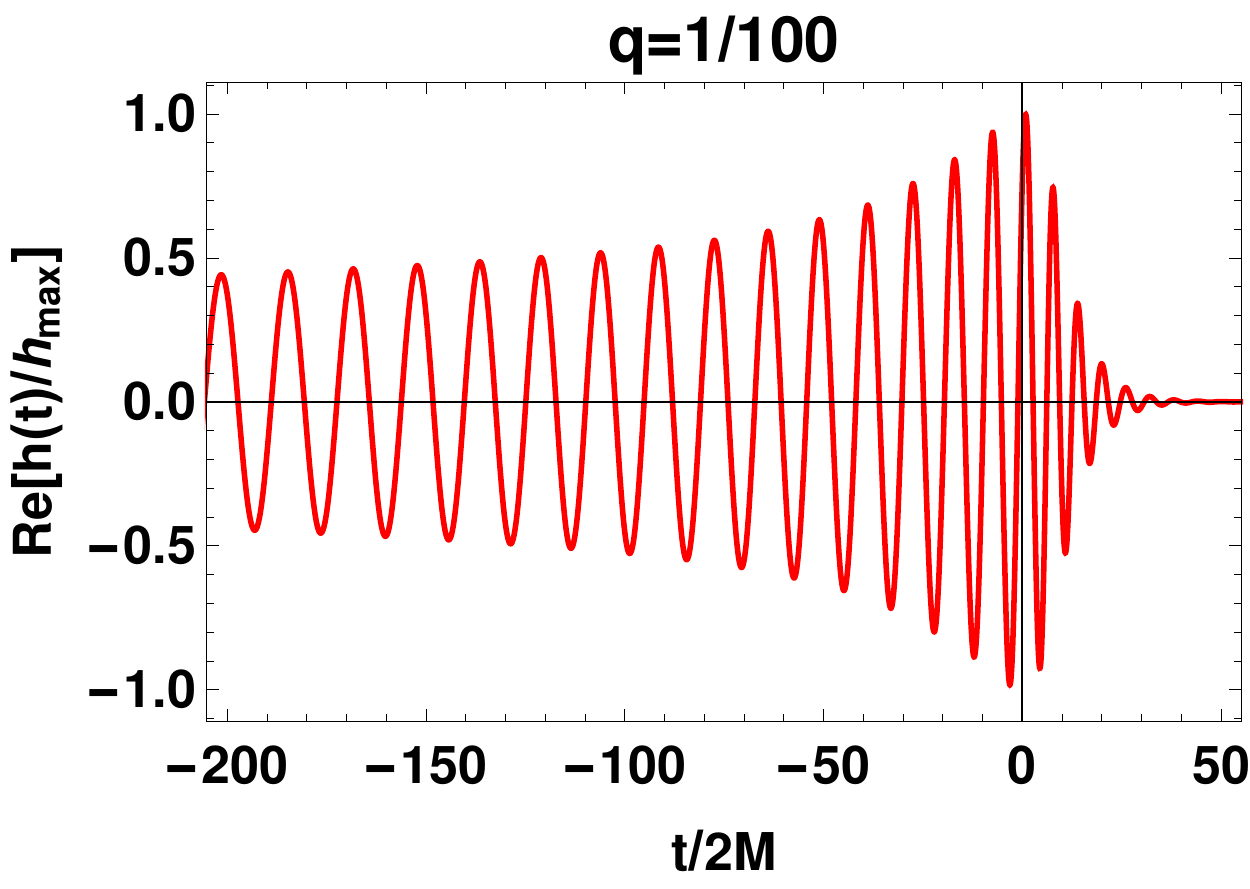}

\caption{Real part of the strain of the GW for different mass ratios $q$, normalized by the peak of its absolute value (at $t=0$). These waveforms were extracted at $r=150M$. The  faster evolution of phase and amplitude found for larger $q$ is due to a faster orbital evolution, see Figure \ref{orbitsa067M}. The echoes appear at later times with smaller amplitudes, as shown in Figure \ref{FewEchoesETH}.}
\label{Strainrealpart}
\end{figure*}

In Figure \ref{Strainrealpart} we show our inspiral-merger-ringdown waveforms. The inspiral phase has twice the orbital frequency displaying the characteristic chirp structure followed by the BH QNM ringdown. In agreement with Figure \ref{orbitsa067M}, the phase and amplitude evolutions are faster for cases with a more massive secondary object. 

\begin{figure*}[ht!]
    \includegraphics[width=0.29\textwidth]{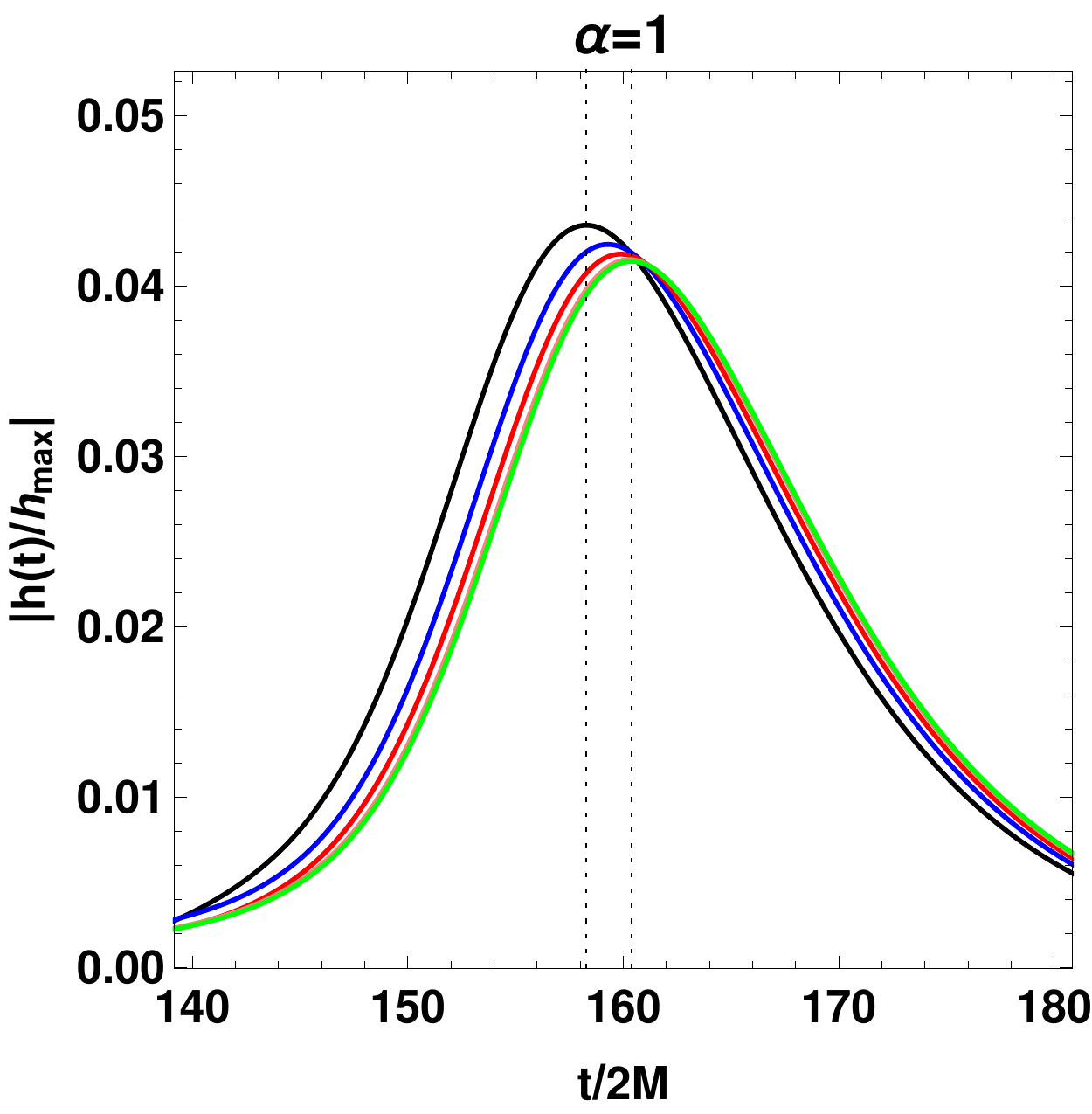}
    \includegraphics[width=0.3\textwidth]{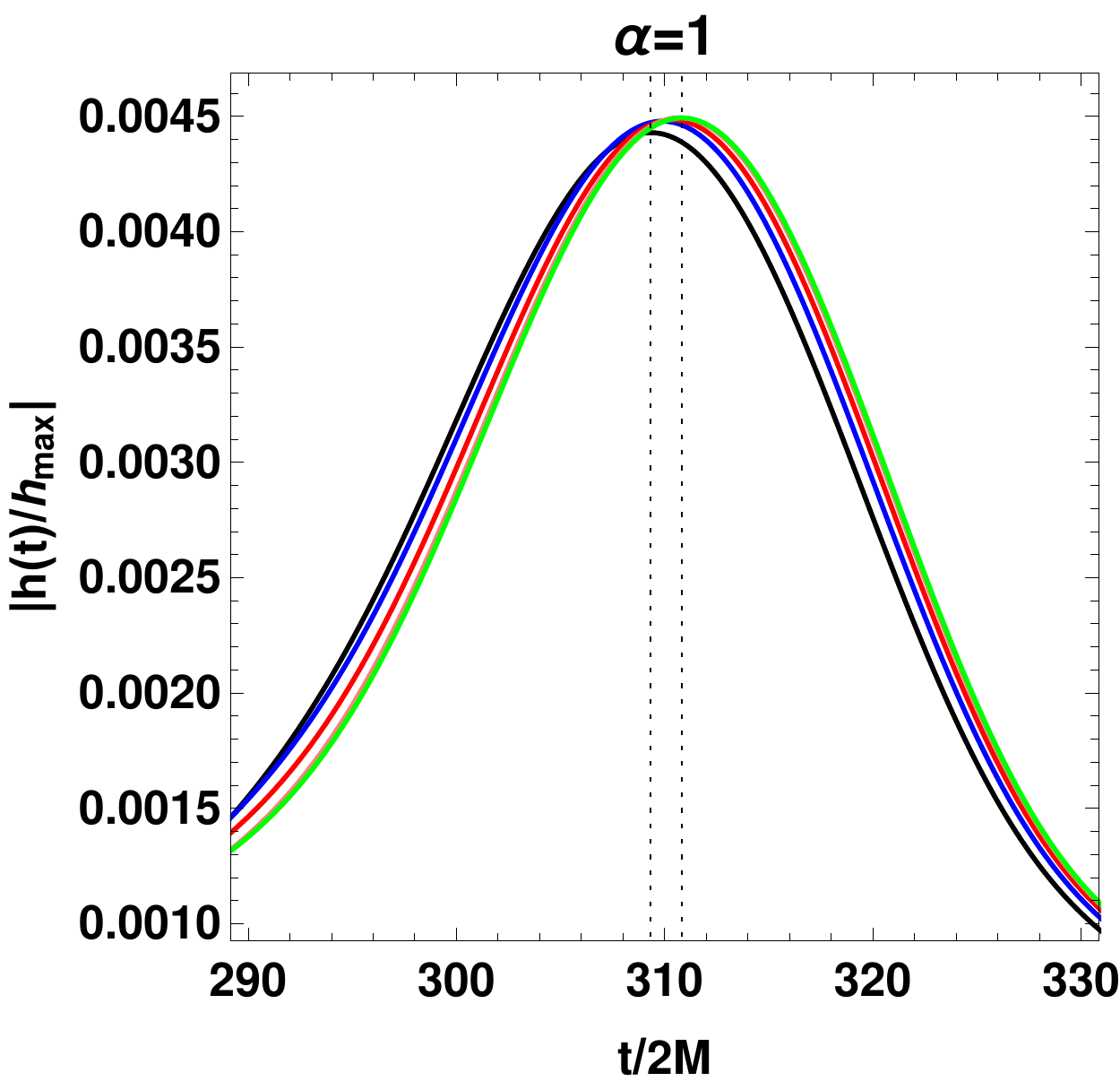}
    \includegraphics[width=0.37\textwidth]{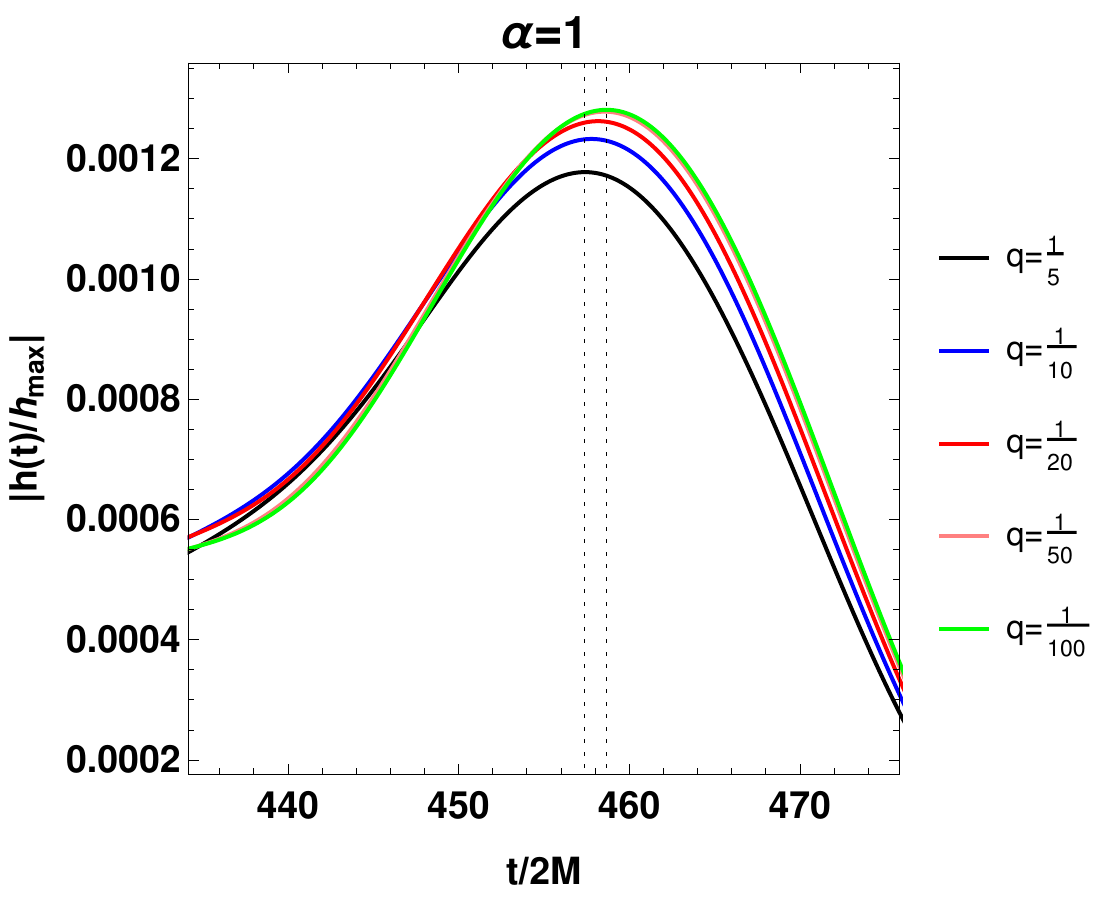}
    \includegraphics[width=0.29\textwidth]{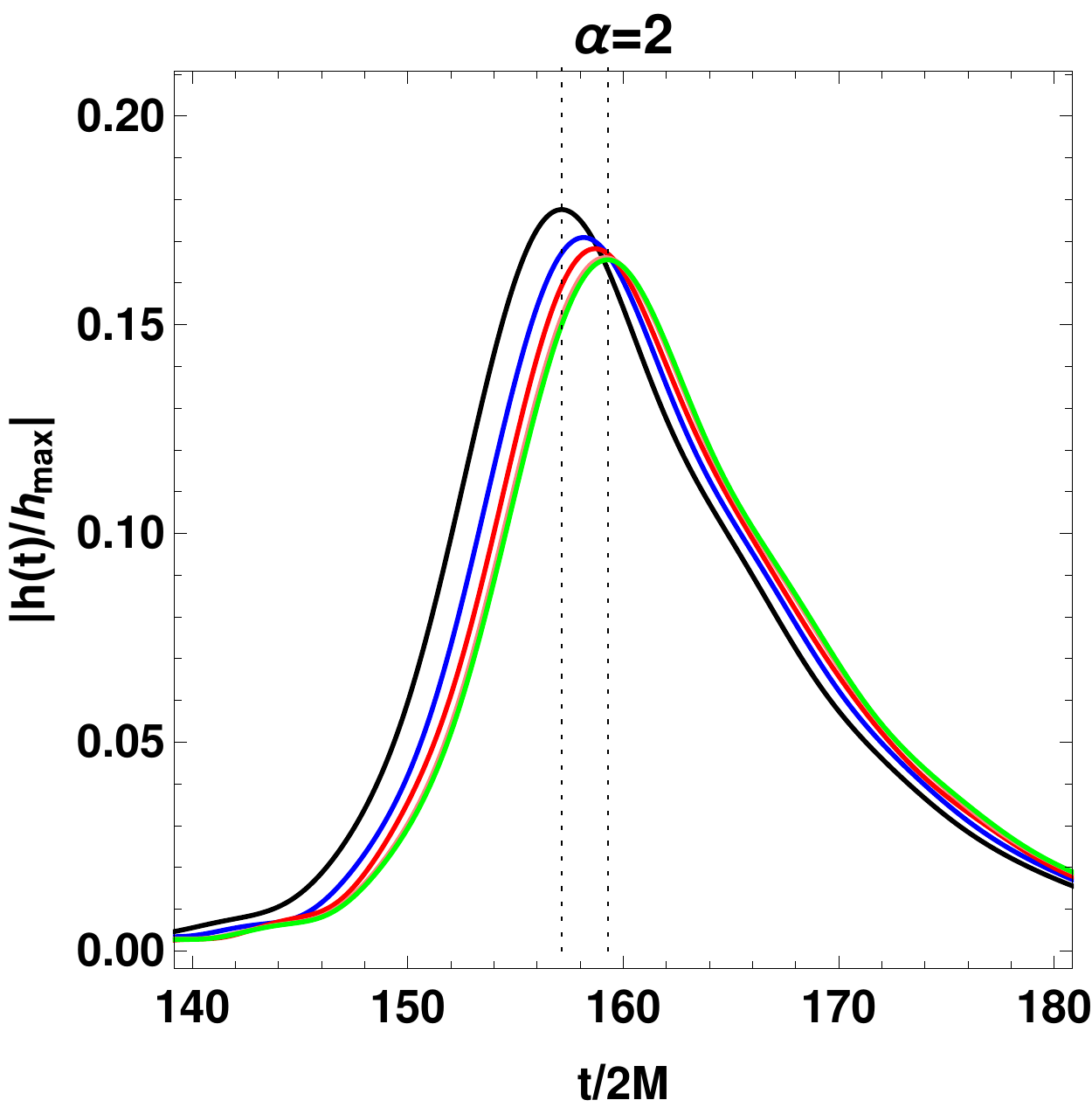}
    \includegraphics[width=0.3\textwidth]{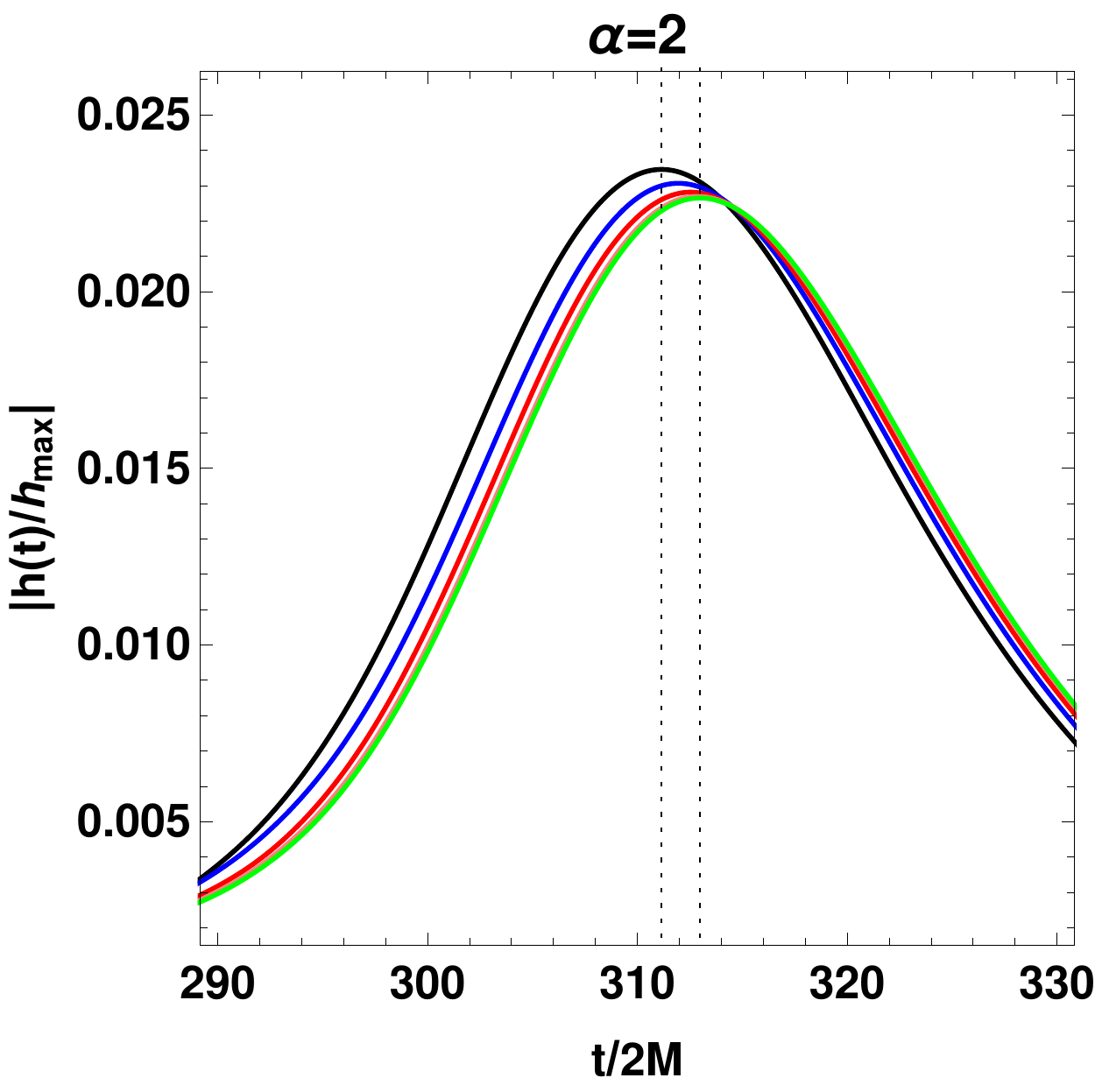}
    \includegraphics[width=0.37\textwidth]{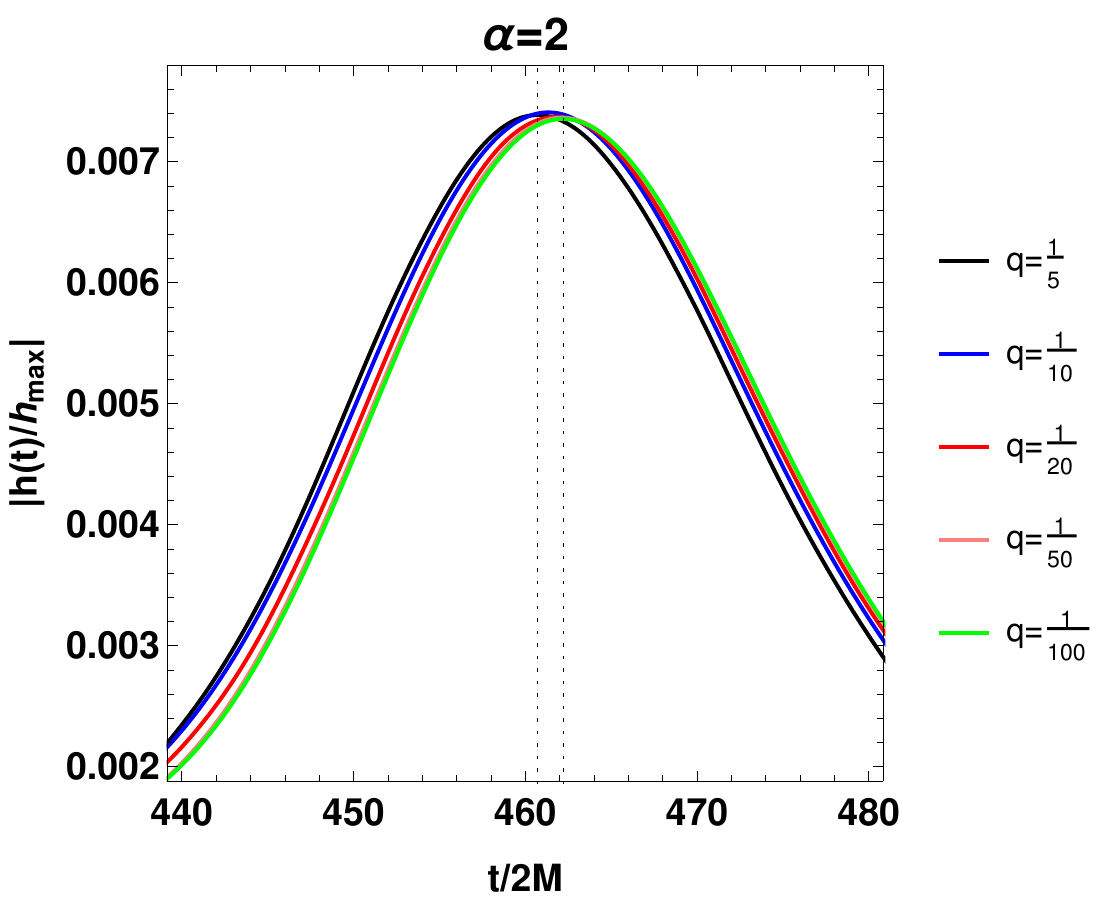}
    
    \caption{Absolute value of the strain of the first three echoes, normalized by the maximum amplitude of the inspiral wave, for $\alpha=1$ and $\alpha=2$ in the top and bottom row, respectively. We also note a later peak time and smaller amplitudes of the first pulse for smaller mass ratios. Each pulse is smaller then the previous one indicating the stability of our model.The vertical dashed lines mark the peak time for $q=1/5$ and $q=1/100$.}
    \label{FewEchoesETH}
\end{figure*}

We analyze the first echoes in Figure \ref{FewEchoesETH}, which shows the absolute value of the first three echoes normalized by the maximum amplitude of the waveform for different choices of $\alpha$. 
The normalized amplitudes increase with increasing values of $q$, independent of the choice of $\alpha$. This result for the normalized amplitude is nontrivial and, coupled with the increase of the overall waveform amplitude for binaries closer to equal masses, could enhance the chances of detection. (The reversed trend for the third echo is discussed in Section \ref{sec:freq} below.)

It can also be seen in Figure \ref{FewEchoesETH} that the first echoes for binaries of larger mass ratios tend to peak earlier, when we set $t = 0$ as the time at which the inspiral-merger-ringdown waveforms shown in Figure \ref{Strainrealpart} have their maximum absolute value. This effect is also related to the orbital evolution shown in Figure \ref{orbitsa067M}. As a heavier infalling particle plunges faster, the position where the particle emits most of the radiation is reached earlier. Therefore, the echoes are expected to peak earlier, as we observe in our waveforms. This trend is reinforced if we suppose that our waveforms would need to go through a rescaling similar to eq.(\ref{rescale}) in order to match future more accurate models from NR. In the rescaled time coordinate $\xi t$, the peak time will appear even earlier for a binary of comparable masses, as $\xi$ decreases with increasing q \cite{Surrogate}.

\begin{figure*}[ht!]
    \centering
    \includegraphics[width=0.4\textwidth]{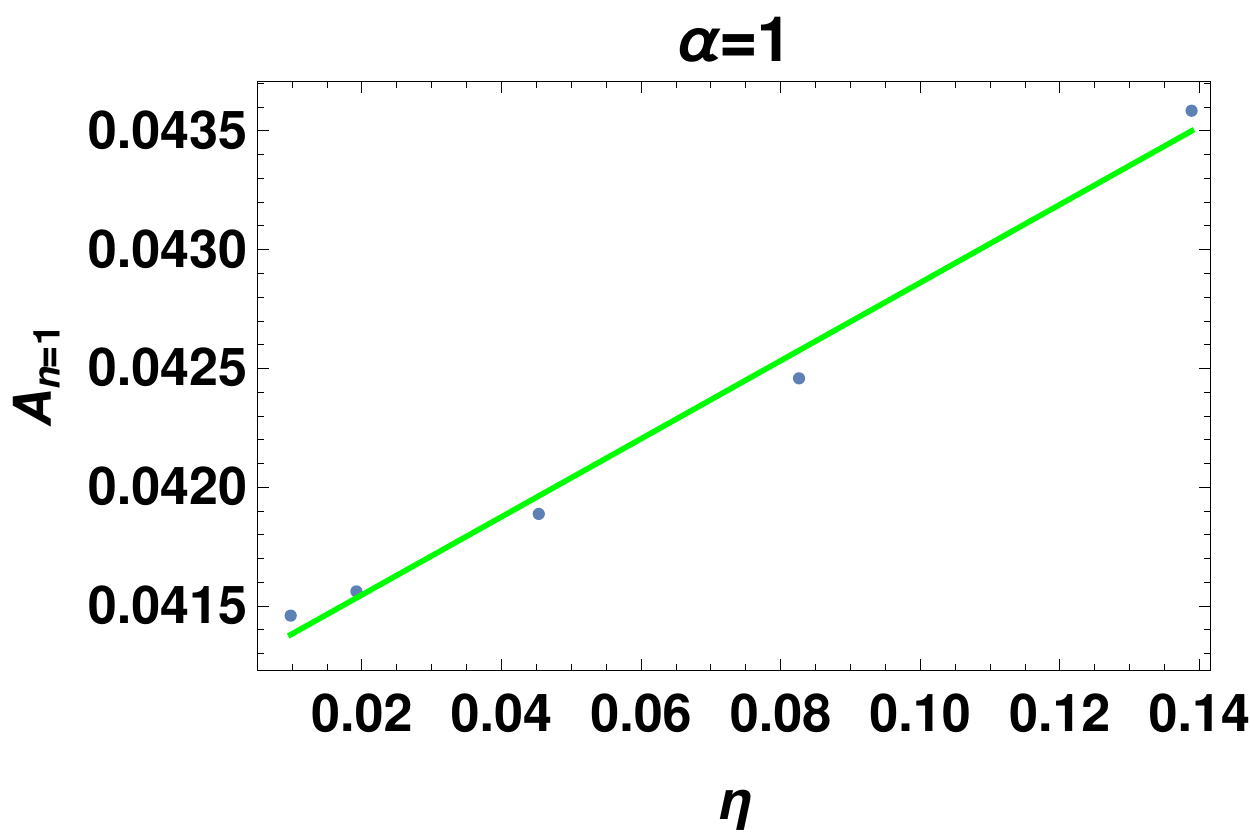}
     \includegraphics[width=0.4\textwidth]{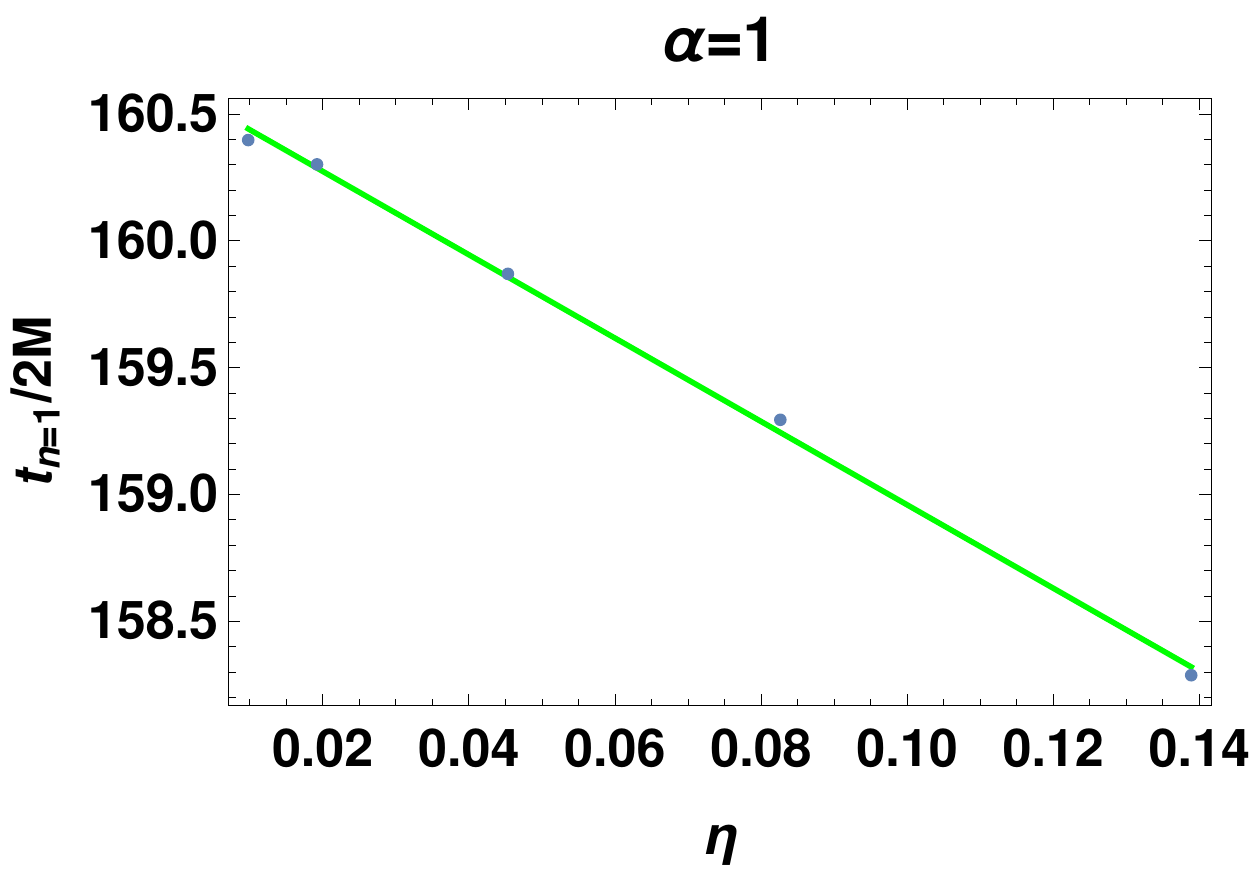}
    \includegraphics[width=0.4\textwidth]{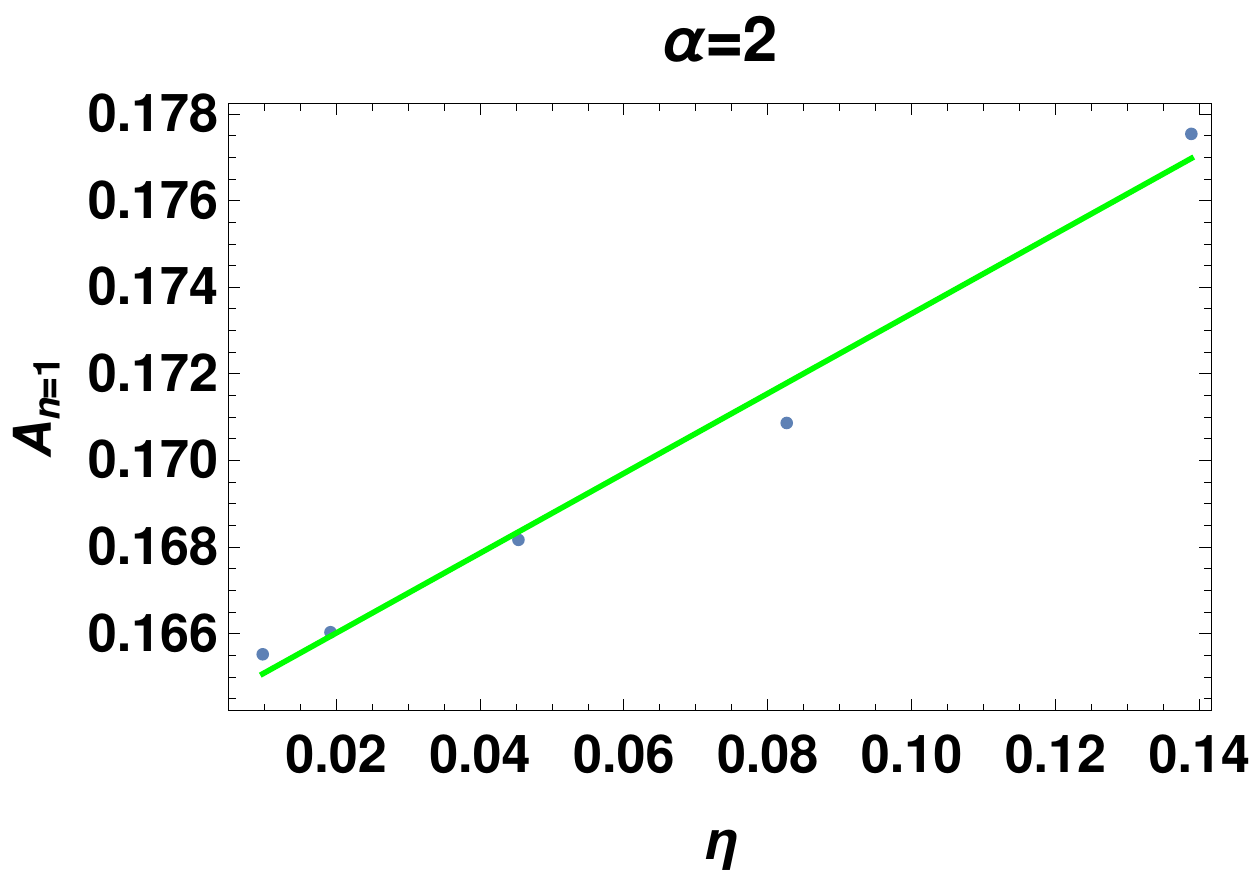}
    \includegraphics[width=0.4\textwidth]{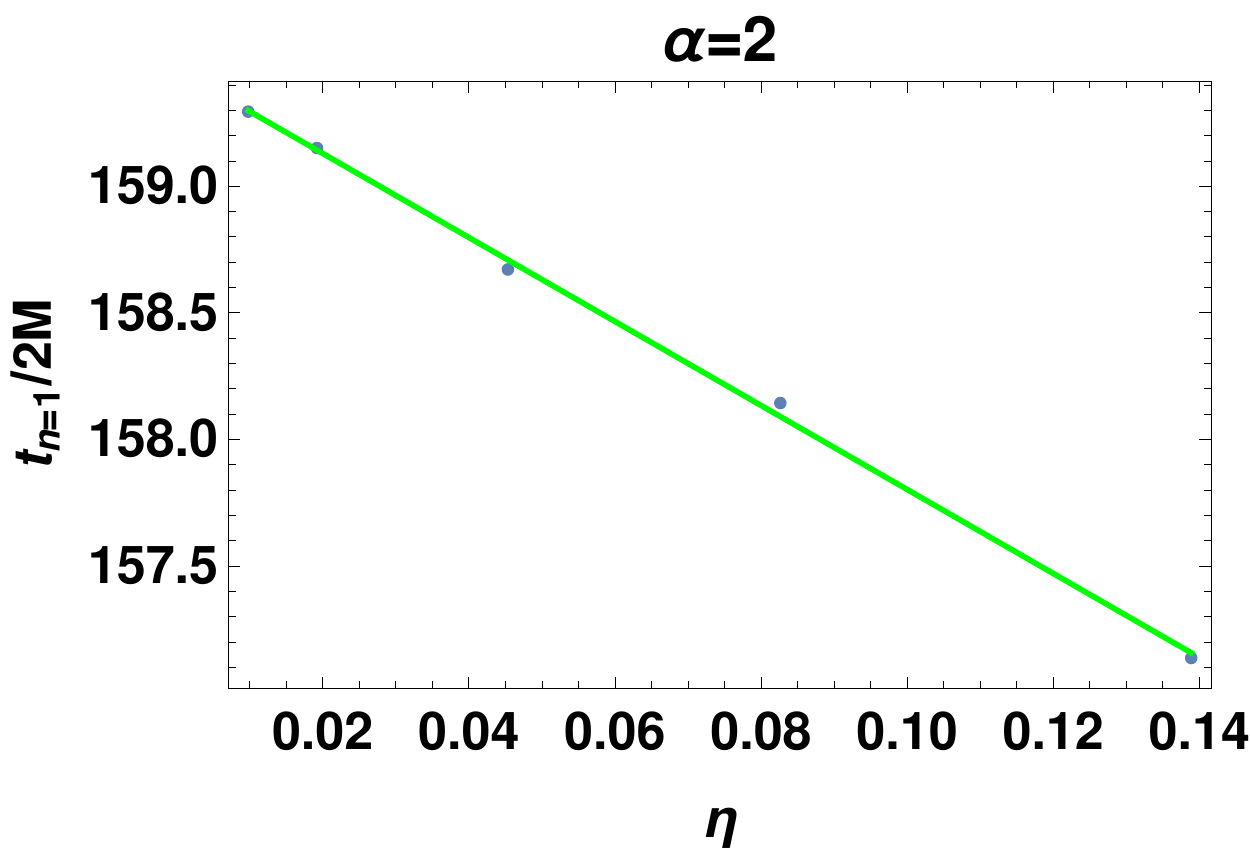}
    \caption{Linear trends for the first echo's  peak amplitude normalized by the maximum amplitude of the inspiral wave (left) and peak time (rigth) as a function of the symmetric mass ratio $\eta$ for $\alpha = 1$ (top) and $\alpha = 2$ (bottom). The peak amplitude depends strongly on the reflectivity parameter $\alpha$ and increases with $\eta$, whereas the peak time is nearly independent of $\alpha$ and decreases with $\eta$ (see main text). The fit paramenters are given in Table \ref{Tablefits}.} 
    \label{AmplitudesETH}
\end{figure*}

To quantify these trends for the first echo peak amplitude and peak time with respect to the (symmetric) mass ratio, we show the linear fits in Figure \ref{AmplitudesETH}, with coefficients provided in Table \ref{Tablefits}. Qualitatively similar results hold for the second and third echoes. The peak amplitudes are strongly dependent on $\alpha$: larger reflectivity leads to higher amplitudes, as expected. In contrast, the slope of the fits for the peak time is nearly independent of $\alpha$. Therefore we can conclude that this time delay is not an effect due to the surface reflectivity, but solely caused by different orbital motions.

\begin{table}[H]\label{Tablefit}
\centering

 \begin{tabular}{ c  c  c} 
 \hline
  \hline
  & $\alpha=1$ & $\alpha=2$  \\ [0.5ex] 
 \hline
 $A_{n=1}$ & $0.0412 + 0.0164 \eta $ & $0.164 + 0.0922 \eta$  \\
 
 $\dfrac{t_{n=1}}{2M}$ & $161-16.4 \eta$  & $160-16.6 \eta$   \\ [1ex] 
 \hline
 \hline
\end{tabular}
\caption{Best fits for the normalized peak amplitude of the first echo ($A_{n=1}$) and the peak time ($t_{n=1}$) for $\tilde{a} = 0.67$ for different values of $\alpha$. Note the weak (strong) dependence of the slope of the peak time (relative amplitude) on $\alpha$.}
\label{Tablefits}
\end{table}

The peak time of the first echo is expected to be a good approximation to the time between consecutive echoes, $\Delta t_{echo}$. This time interval can be estimated naively as $\Delta t_{echo}\approx 2|r_{*}^{0}|$ and it appears to coincide with the BH scrambling time (which, in the context of quantum information, is understood as the time it takes to recover information previously thrown into a BH \cite{Scramblingtime,Sekino_2008}). However, our results for $t_{n=1}$ support that  $\Delta t_{echo}\approx 2|r_{*}^{0}| + \mathcal{O}(q)$ is a better approximation. Neglecting this $q$ dependence would lead to an error in the position of the ECO wall $r_{*}^{0}$. Assuming that the underlying quantum gravity corrections in the ECO spacetime appear at the Planck length above the horizon \cite{AfshordiDetection}, we have
\begin{equation}
 \int^{r_{+}(1+\epsilon)}_{r_{+}} \sqrt{g_{rr}}dr \sim \ell_{Planck}.
\end{equation}
Therefore, returning to SI units, we have:
\begin{equation}
 \delta r \equiv r_{+}\epsilon = \dfrac{\sqrt{1-\tilde{a}^{2}}\ell_{Planck}^{2}c^{2}}{4GM(1+\sqrt{1-\tilde{a}^{2}})},
\end{equation}
which for $a=0.67M$ and $M=30M_{\astrosun}$ is $\delta r\approx 6\times10^{-76}$ meters and, according to eq. (\ref{rstar}), gives $r_{*}^{0}\approx -434 M$. Figure \ref{AmplitudesETH} shows that the variation of the first echo's peak time can be as large as $4M$. This could result in an error of $2M$ in the determination of $r_{*}^{0}$. Therefore $r_{*}^{0}$ could be either determined as $-434M$ or $-432M$, for example. Translating to the usual $r$ coordinate, the difference between these positions would be $|\delta r - \delta r_{\rm naive}| \approx 8\times10^{-76}$ meters (where $\delta r_{\rm naive}$ is the estimate for $\delta r$ when not including the effects discussed in this work). This implies that ignoring the time-delay dependence on the progenitor properties can lead to 130\% error in inferring the position of the reflective wall ($|\delta r - \delta r_{\rm naive}| \approx \delta r $).

\subsection{Frequency dependence}\label{sec:freq}

\begin{figure}
\centering
  \includegraphics[width=0.45\textwidth]{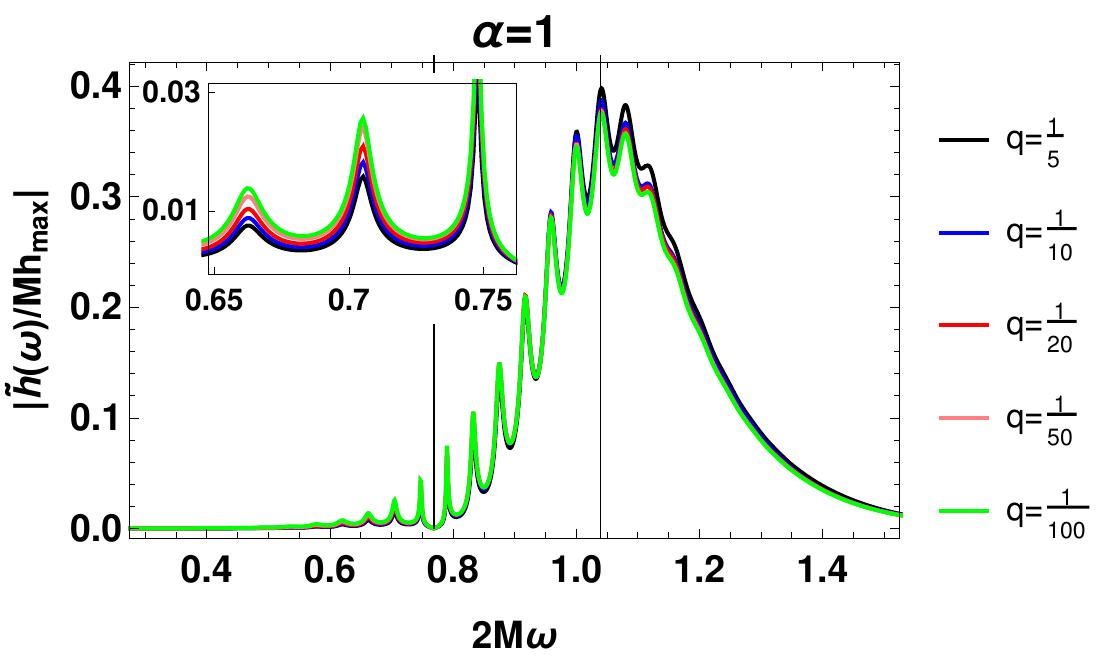}
   \includegraphics[width=0.45\textwidth]{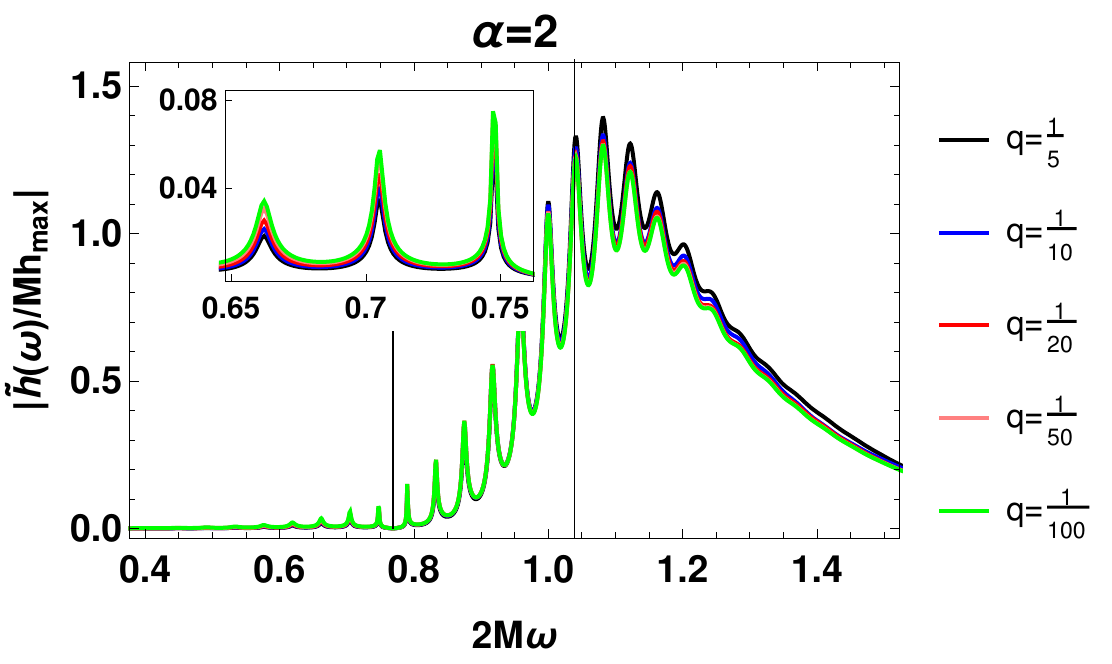}
  \caption{Absolute value of the strain of the full echo waveform in the frequency space ($\tilde{h}_{(\omega)}^{echoes}$) as extracted at $r=150M$ and normalized by the peak of the inspiral waveform. The two vertical lines represent $\omega_{SR}$ and the fundamental BH QNM (from left to right). The insets zoom in on the $q$-dependence at low frequencies. We see that the amplitude slightly increases (decreases) with $q$ at high (low) frequencies.}
\label{FirstEchoeETHfrequencyspace}
\end{figure}

 Figure \ref{FirstEchoeETHfrequencyspace} shows the amplitude of the echoes in the frequency domain for different mass ratios, $q$. The peak frequency of the echoes matches that of the BH fundamental quasinormal mode solid vertical lines, independent of $q$. 
 However, at lower frequencies, and especially below the superradiance frequency $2M\Omega_H$, the resonance structure due to the quasi-periodicity of the echoes becomes more prominent. The resonance peaks can be identified with the $(2,2,n)$ QNM overtone frequencies of the BH potential \emph{and} of the ECO in our model. Being characteristic frequencies, their detection would provide information to rule out or support different ECO models.
 
 Moreover, even though larger $q$ leads to more power at high frequencies, the trend reverses at lower frequencies, where we have louder echoes for smaller mass ratios. This also influences later echoes, which have lower frequencies. This is the reason why the third echo in Figure \ref{FewEchoesETH}) shows larger amplitudes for smaller mass ratios.

\subsection{Echo dependence on BH spin}\label{spindependence}

It was argued in \cite{Afshordi1} that ECOs motivated by quantum deviations from a Kerr BH should have their reflective surface located at the same  \textit{proper distance} (of order $\ell_{Planck}$) from the ``would be" horizon for all objects. This means that, within the same theory, ECOs of different masses and spins will have surfaces at a different positions $r_{*}^{0}$. 

In \cite{Scramblingtime} it was proposed that the following relation should be satisfied for two ECOs with spin parameters $a_{1}$ and $a_{2}$:
\begin{eqnarray}\label{rstarzero}
  \left.(T_{H} r_{*}^{0})\right|_{a_{1}}  \approx \left.(T_{H} r_{*}^{0})\right|_{a_{2}}, 
\end{eqnarray}
where the BH temperature $T_{H}$, given by (\ref{temp}), and the position of the wall $r_{*}^{0}$  are functions of the spin parameter $a$. Therefore we expect a larger time delay between consecutive echoes for highly spinning ECOs. Given our previous choice of $r_{*}^{0}=-150M$ for $a=0.67M$, the corresponding position for the reflective wall of an $a=0.9M$ ECO is found to be $r_{*}^{0}\approx-210.5M$.
  
\begin{figure}[ht!]
      \centering
       \hspace{0.4cm}\includegraphics[width=0.51\textwidth]{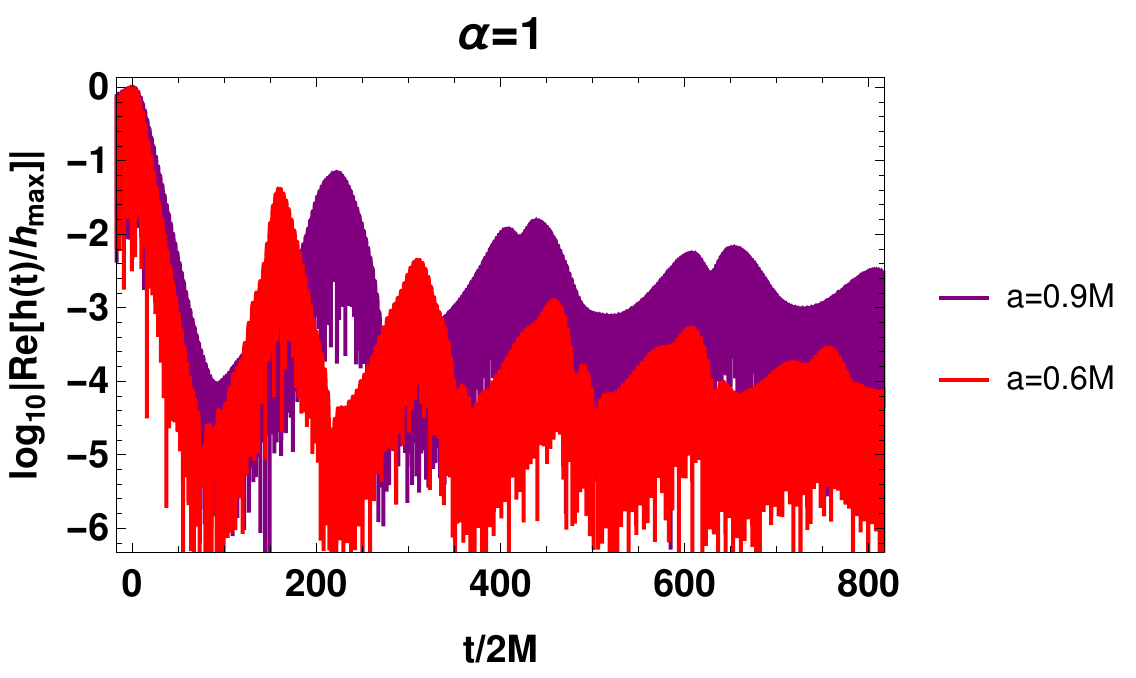}
      \hspace{0.4cm}\includegraphics[width=0.51\textwidth]{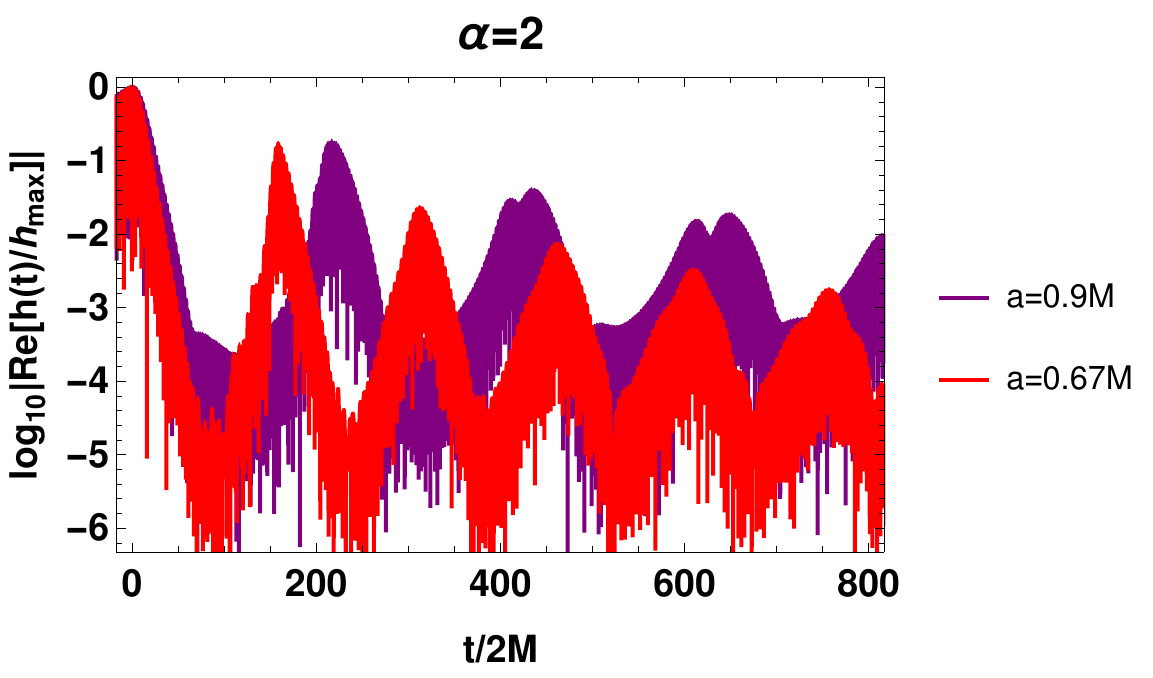}
      \caption{Real part of the strain of the ringdown after the merger and up to the first few echoes for different spins (and corresponding positions of the reflective wall) and fixed mass ratio ($q=1/20$), for two values of $\alpha$. The rapidly rotating ECO shows echoes with larger amplitude and with a double peak (see main text).}
      \label{DifspinETH}
  \end{figure}

 We now apply our method for the choice of parameters: $a=0.9M$, $q=1/20$, $r_{*}^{0}=-210.5M$, and compare the results with the $a=0.67M$ case (see Figure \ref{DifspinETH}). Two main differences can be noticed between the two cases. First, consecutive echoes take longer to appear for larger spins, as previously expected and in agreement with \cite{Afshordi2, PaniAnalytical}. Second, the amplitude of the echoes is enhanced for $a=0.9M$. Since $T_{H}\rightarrow 0$ in the extremal limit (i.e. $a \to M$), the Boltzmann reflectivity becomes sharper and a smaller range of frequencies (therefore less energy) should be reflected; in contrast, the  superradiant amplification grows in the extremal limit \cite{Micchi2}. Therefore, the effect of increasing spin on the echo amplitudes is not monotonic. For $a = 0.9M$, the superradiant amplification appears to dominate over the Boltzmann suppression, leading to larger echoes. 

 We can also notice a double peak structure for later echoes in the $a=0.9M$ case. As seen in \cite{Micchi1}\footnote{In \cite{Micchi1}, the observed double peak structure comes from contributions with positive and negative frequencies, and the contribution from superradiant frequencies is negligible in the scalar case. Here the negative frequency contribution is eliminated by our choice of reflectivity, whereas the superradiant contribution is much stronger in the gravitational case.}, we attribute this characteristic to the double hump structure of the transfer function, which leads to an echo structure described by two different sine-gaussian wave packets. This effect is stronger in the larger spin case because the superradiant contribution becomes more prominent, see Figure \ref{Ktransferfunctions}. The double peak structure reinforces the proposal made in \cite{Micchi1} of a generalization of the echo template found in \cite{Cardosomorphology} to a train of double sine-gaussian packages.

\subsection{Resonances}\label{Resonances}

 We also investigate whether the orbital phase of the waveform could be modified by resonances of the QNMs of the ECO. Therefore, we look for relative differences between the orbital phase of the ECO waveform, as obtained by eq. (\ref{final}), and the expected BH waveform, given by the first term of (\ref{final}).
 
We report that we do not find any measurable modification of the waveform in the inspiral phase. This result agrees with the analytical work performed in \cite{CardosoRessonances}  where it was found that the resonances are crossed very quickly during the inspiral phase of the motion; therefore their impact on this stage of the GW emission is negligible. Analysing Figure \ref{FirstEchoeETHfrequencyspace}, one could have anticipated that this effect would be negligible: small frequencies ($\omega<\omega_{SR}$), emitted during the orbital phase, are highly suppressed due to the small reflectivity.

In \cite{Tidal}, it was found that the orbital phase of the inspiral waveform emitted by an ECO would have a different evolution from the one emitted by a BH. While this effect could significantly impact the inspiral phase over several orbits, we do not expect it to have a large effect on the echo waveforms in our study. 

\section{Detectability}
\label{sec:detect}

We are now able to assess the detectability of the  echoes from ECOs with current and future gravitational wave observatories. The expected signal-to-noise ratio (SNR) of an event is defined as:

\begin{equation}\label{SNR}
\left(\dfrac{S}{N}\right)^{2} = 4 \bigintsss_{0}^{\infty}df \dfrac{|h(f)|^2}{S_{n}(f)}, 
\end{equation}
where $S_{n}(f)$ is the  single-sided noise spectral density of the detector and has units of $\text{Hz}^{-1}$, and $f=\omega/2\pi$ as usual \cite{Maggiorebook}. Additionally, we take the average over the sky of the detector response function (see Table 7.1 in \cite{Maggiorebook}).

\begin{figure*}
    \centering
\hspace{-2.8cm}\includegraphics[width=0.32\textwidth]{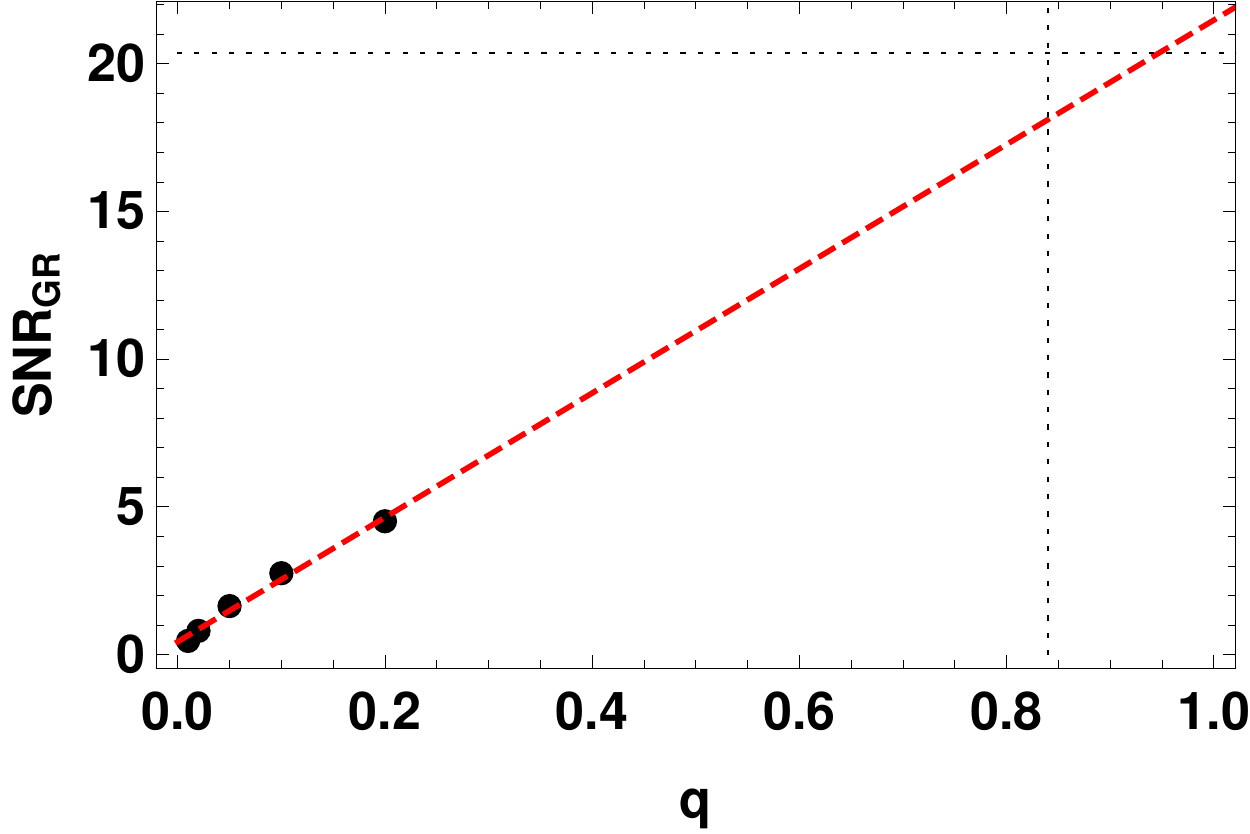}
\hspace{0.8cm}\includegraphics[width=0.32\textwidth]{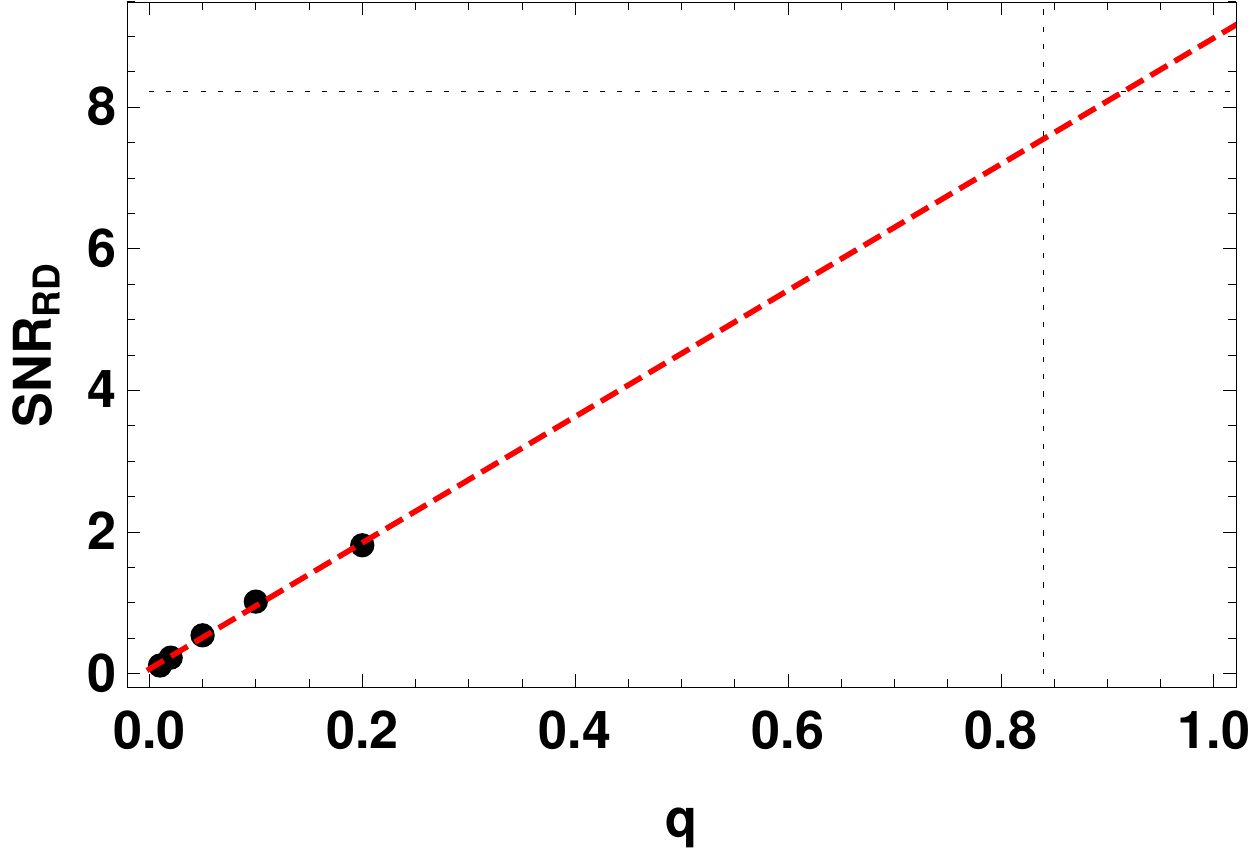}
\hspace{2.5cm} \includegraphics[width=0.32\textwidth]{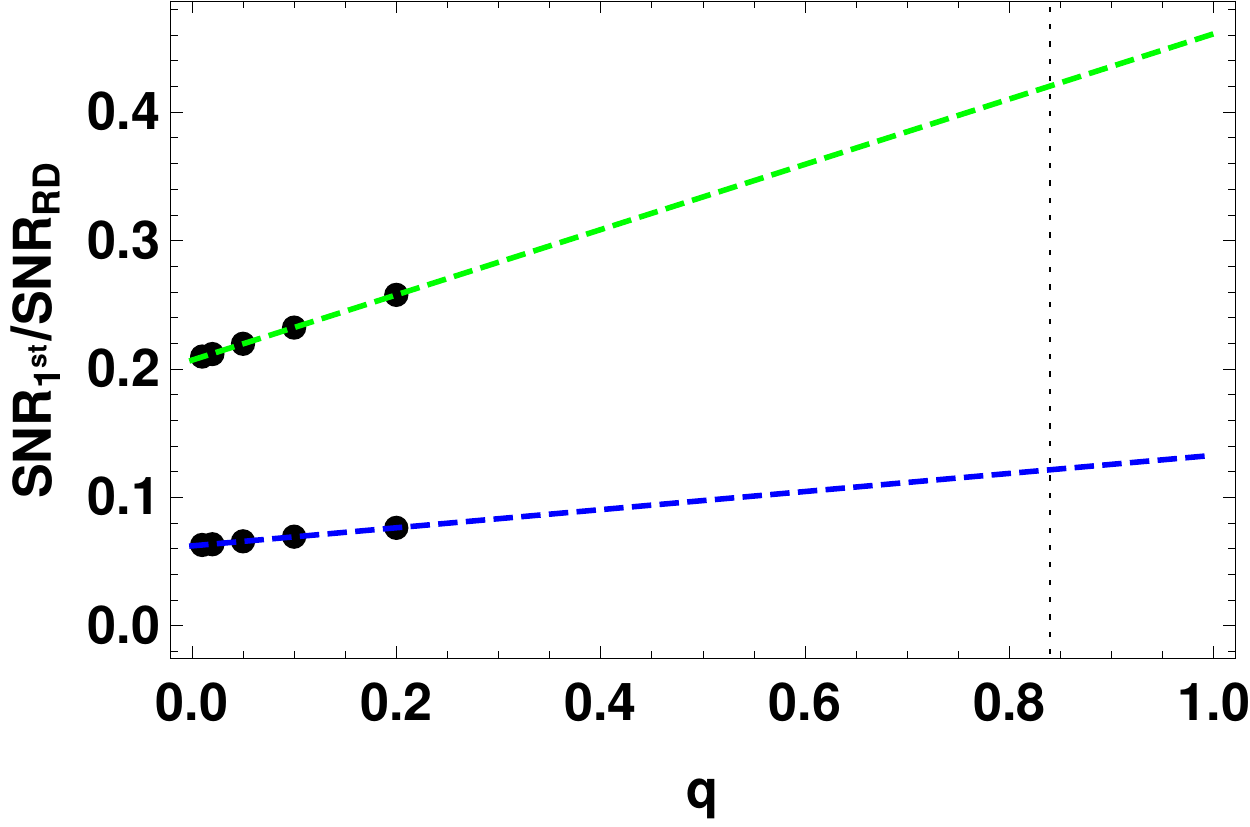}
\hspace{0.cm}\includegraphics[width=0.53\textwidth]{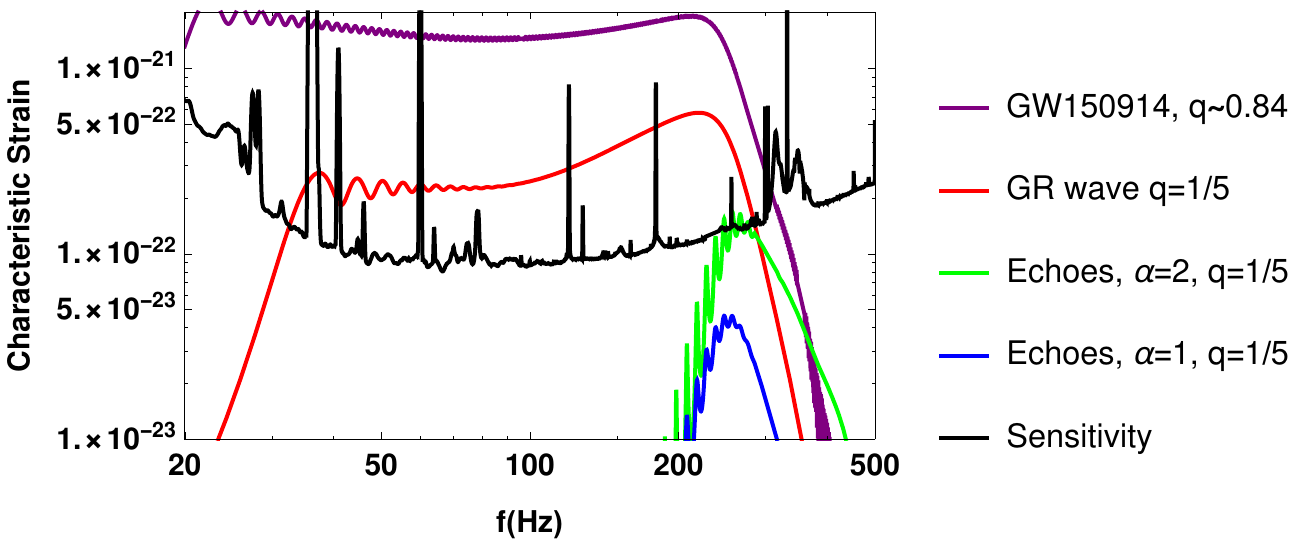}

    \caption{Top: single detector (LIGO Hanford - O1) SNR of the  inspiral-merger-ringdown wave predicted by GR (left) and the corresponding ringdown SNR (right) as a function of the mass ratio $q$. The cases presented correspond to the waveforms shown in Figure \ref{Strainrealpart}, with the remnant mass and distance to the source scaled to GW150914.  The black dashed lines show the mass ratio and SNR values for GW150914, in good agreement with our linear extrapolation. Bottom: single detector SNR of the first echo, normalized by the ringdown SNR for $\alpha = 1, 2$ (left) and characteristic strain examples (right). }
    \label{XX}
\end{figure*}

In Figure \ref{XX} we fix the final mass of the merger and the distance to the source to those of GW150914 as a representative example (with the parameters given in \cite{ImprovedGW150914}) and we use the LIGO Hanford detector noise curve during O1  \cite{LIGOGWOSC,LIGOASD}. On the top left panel we show the single detector SNR of the full inspiral-merger-ringdown predited by GR (${\rm SNR}_{\rm GR}$) obtained with our setup for mass ratios from 1:100 to 1:5 (see Figure \ref{Strainrealpart}). The dotted lines indicate the observed mass ratio and single detector SNR for GW150914, in good agreement with the linear extrapolation of our results. The top right panel presents similar results, this time restricted for the ringdown SNR (${\rm SNR}_{\rm RD}$), assumed to start 3 ms after the peak of the gravitational wave at merger. In the bottom left panel we have the expected range for the single detector SNR for the first echo (${\rm SNR}_{1^{\rm st}}$), normalized by the corresponding ${\rm SNR}_{\rm RD}$. In the bottom right panel we show some examples of the characteristic strain together with the detector noise curve \cite{Moore_2014}.

Assuming ${\rm SNR}_{1^{\rm st}} > 8$ for detection of the first echo would require ${\rm SNR}_{\rm RD} \gtrsim 19$ (in the most optimistic case with $\alpha = 2$) and ${\rm SNR}_{\rm RD} \gtrsim 66$ (in the most pessimistic case with $\alpha = 1$). These values are $\sim 2-8$ times stronger than the observation of GW150914 during O1. So, what are our realistic chances of detecting echoes?

If GW150914 had been detected during O3, its SNR would almost be twice as large as it was in O1. This threshold should be crossed in O4, currently scheduled to start in the Fall of 2021 \cite{LIGOnext}. The rate of observations (triggers) during O3 was approximately one event per week. During O4 (scheduled to have 18 months of observations), the currently expected improvements in sensitivity should provide a total of approximately 260 events. As ${\rm SNR} \propto {\rm (distance)}^{-1}$ and the expected number of detections $n$ is proportional to ${\rm (distance)^3}$, we can estimate the number of events detected with an SNR higher than a certain threshold to be
\begin{equation}
    n \sim 260\left(\frac{{\rm SNR}_{\rm thresh}}{{\rm SNR}_{\rm min}}\right)^{-3}\,,
\end{equation}
where ${\rm SNR}_{\rm min}$ is the minimum SNR for the detection of the event, which we set to 8. This estimate yields approximately 1-2 events/year during O4 with ${\rm SNR}_{\rm thresh} \sim 40$ (twice as loud as was observed for GW150914), which is the required SNR for the detection of the first echo in the most optimistic case with $\alpha = 2$. In the most pessimistic case with $\alpha = 1$, we obtain an estimate of 0.03 events/year. Since our SNR estimates depend on the linear extrapolation from $q\ll 1$ to $q \sim 1$, they could be off by a factor of 2-3. However, we expect an improvement by one or two orders of magnitude in sensitivity for future 3G detectors (such as the Einstein Telescope and the Cosmic Explorer) \cite{3rdgenDetectors}. Therefore, our qualitative conclusions about the detectability of echos should not be overly sensitive to the nature of extrapolation.

\section{Conclusions} \label{conclusions}
 
Here we investigate, for the first time, the detectability of echoes from ECOs as predicted by a realistic model of orbital excitation and ECO reflectivity. Going beyond the pure geodesic plunge description for the inspiral, we were able to determine the effects of the mass ratio on the echoes. 

The first effect is the enhancement of the amplitude of the first echo (normalized by the peak of the inspiral waveform) with increasing mass ratio. The second effect is a shift of the peak time of the first echo. If echoes do exist, the first echo detections will most likely be able to reconstruct only the first echo. In this  scenario, the compactness of the ECO will be determined by peak time of the first echo alone. The non-inclusion of mass ratio corrections can lead to a  wrong estimate for the reflective wall's position.
 
Our model predicts a \emph{range} for the reflectivity of the ECO, rather than simply setting an upper limit. Therefore this model is directly testable by observations. We expect to start probing the most optimistic projections for the detection of the first echo in the next LIGO observing run (O4), currently scheduled to start in the Fall of 2022. It is also possible that the stacking of LIGO events could improve the chances for an early detection. The most pessimistic scenarios will be easily probed in the 2030s with 3G ground based detectors and, of course, with LISA.

\begin{acknowledgments}
We thank  Naritaka Oshita, Qingwen Wang, Jahed Abedi, Ramit Dey, Cole Miller, Mauricio Richartz and Sayak Datta for useful discussions and comments. LFLM was supported in part by grant 2017/24919-4  of the S\~ao Paulo Research Foundation (FAPESP) and by Coordena\c c\~ao de Aperfei\c coamento de Pessoal de N\'ivel Superior - Brasil (Capes) - Finance code 001 through the Capes-PrInt program. NA acknowledges support from the University of Waterloo, the Natural Sciences and Engineering Research Council of Canada, and the Perimeter Institute for Theoretical Physics. CC acknowledges support from grant 303750/2017-0 of the Brazilian National Council for Scientific and Technological Development (CNPq), from the Simons Foundation through the Simons Foundation Emmy Noether Fellows Program at Perimeter Institute and by NASA under award number 80GSFC17M0002. We are grateful for the hospitality of Perimeter Institute where most of this work was carried out. 
Research at Perimeter Institute is supported in part by the government
of Canada through the Department of Innovation, Science and Economic
Development and by the Province of Ontario through the Ministry
of Colleges and Universities.

\end{acknowledgments}

\appendix
\section{Sasaki-Nakamura Formalism}\label{appendixA}
\subsection{SN-Teukolsky transformations}\label{appendixtransform}

In Section \ref{Setup} we extensively use the relations between the SN asymptotic amplitudes (given in (\ref{eqinSN}) and (\ref{equpSN})), and the corresponding Teukolsky asymptotic amplitudes (given in (\ref{eqin}) and (\ref{equp})). Here we summarize these relations.

In our notation, the SN-Teukolsky relations can be written as
\begin{subequations}\label{A1}
\begin{align}
    {}_{SN}B^{\rm trans}_{lm\omega}&= d B^{\rm trans}_{lm\omega}, \label{inrealtionsa} \\
{}_{SN}B^{\rm ref}_{lm\omega}&=- \dfrac{c_{0}}{4 \omega^{2}}B^{\rm ref}_{lm\omega} \label{inrealtionsb}, \\
    {}_{SN}B^{\rm inc}_{lm\omega}&=-4\omega^{2} B^{\rm inc}_{lm\omega}, \label{inrealtionsc}
\end{align}
\end{subequations}
for the in-mode, whereas for the up-mode we have 
\begin{subequations}\label{A2}
\begin{align}
    {}_{SN}C^{\rm trans}_{lm\omega}&=-\dfrac{c_{0}}{4\omega^{2}}C^{\rm trans}_{lm\omega}, \label{uprealtionsa} \\
{}_{SN}C^{\rm ref}_{lm\omega}&=d C^{\rm ref}_{lm\omega}, \label{uprealtionsb}\\
    {}_{SN}C^{\rm inc}_{lm\omega}&= g C^{\rm inc}_{lm\omega}.\label{uprealtionsc}
\end{align}
\end{subequations}

In equations (\ref{A1}) and (\ref{A2}), the following definitions are used:

\begin{subequations}
\begin{align}\label{constantsd} 
c_{0}&= {}_{-2}\lambda_{lmc} (2 + {}_{-2}\lambda_{lmc}) \nonumber \\ & \quad \quad \quad- 12\omega (-a m + M\textit{i} + a^{2} \omega),\\
d&= -2  \sqrt{2Mr_{+}} (2 a m + \textit{i} (r_{-}-r_{+}) - 4M \omega r_{+} )\nonumber \times\\&\quad \quad \quad (a m + \textit{i} (r_{-}-r_{+}) - 2M\omega r_{+}),\\
b_{0}&={}_{-2}\lambda_{lm\omega}^{2}+2{}_{-2}\lambda_{lm\omega}-96k^2M^2+72k M r_{+}\omega \nonumber\\
& \quad -12r_{+}^2\omega^{2} -i(16kM({}_{-2}\lambda_{lm\omega}+3-3Mr_{+}^{-1}) \nonumber \\ &\quad\quad\quad\quad\quad\quad\quad-12M\omega-8{}_{-2}\lambda_{lm\omega}r_{+}\omega ),  \\
 g&=\dfrac{-b_{0}}{4k(2Mr_{+})^{3/2}(k+2i(r_{+}-M)(4Mr_{+})^{-1})}.
\end{align}
\end{subequations}

The quantities $c_{0}$ and $d$ can be found in earlier works \cite{SasakiOriginal,SasakiReview}, but $g$ was first derived in \cite{Holdomnewwindows}. We independently derived an equivalent expression for $g$, which is however much more involved and we do not reproduce it here. Therefore we use $g$ as given in equation (\ref{constantsd}) in the flux formulas found in the next Section.

\subsection{Fluxes}\label{appendixfluxes}

In Section \ref{Setup}, we discuss the boundary condition for the Green's function. In our model, we impose that the ingoing and outgoing fluxes at the reflective wall of the homogeneous solution $X_{lm\omega}^{ECO}$ should be proportional. In order to construct the necessary fluxes from the asymptotic behavior of $X_{lm\omega}^{ECO}$, we use expressions found in \cite{Tagoshiechoesandhowls,Brito:2015oca,Holdomnewwindows}. 

A general solution $R(r)$ of the $s=-2$ radial Teukolsky equation behaves, near $r\rightarrow r_{+}$, as :

\begin{equation}\label{eqgen}
    R \sim A^{\rm in}\Delta^{2} e^{-\textit{i} k r_{*}} + A^{\rm out}e^{\textit{i} k r_{*}},
\end{equation}
which is equivalent, via the transformations (\ref{A1}) and (\ref{A2}), to a solution of the SN equation of the form:
\begin{equation}\label{eqgenSN}
    X \sim _{SN}A^{\rm in} e^{-\textit{i} k r_{*}} + _{SN}A^{\rm out}e^{\textit{i} k r_{*}},
\end{equation}
It can be proven that the in and out-going fluxes at the horizon are \cite{Teukolsky3,Brito:2015oca,Tagoshiechoesandhowls}:

\begin{eqnarray}
F_{H}^{\text{in}}&=& \dfrac{128\omega k (2Mr_{+})^ {5}(k^2+4\bar{\epsilon}^2)(k^{2}+16\bar{\epsilon}^{2})}{|C|^{2}}\left|A^{in}\right|^{2}\nonumber \\ &=&\dfrac{8 \omega k}{|C|^{2}}\left|_{SN}A^{in}\right|^{2}, \\
F_{H}^{\text{out}}&=& \dfrac{\omega}{2k(2Mr_{+})^3(k^{2}+\bar{\epsilon}^{2})}\left|A^{out}\right|^{2}\nonumber \\ &=&\dfrac{8 \omega k}{|b_{0}|^{2}}\left|_{SN}A^{out}\right|^{2},
\end{eqnarray}
where the definitions:
\begin{align}
\bar{\epsilon} &= \dfrac{r_{+}-M}{4Mr_{+}}, \\
|C|^{2}&=((_{-2}\lambda_{lmc} + 2)^2 + 4 a m \omega -4 (a \omega)^2)\times  \\ &\quad\quad(_{-2}\lambda_{lmc}^2 +36 m a \omega - 36 (a \omega)^2)\nonumber \\&\quad+ (2 _{-2}\lambda_{lmc} +3) (96 (a \omega)^2 - 48 m a \omega)\nonumber \\&\quad+144 \omega^2(M^2 - a^2), \nonumber
\end{align}

are used. The constant $C$ first appeared in the context of solving the Teukolsky equation \cite{Teukolsky3}.

Comparing equation (\ref{eqgenSN}) to (\ref{Xeco}), it is straightforward to see that the fluxes for our ECO model are given by:

\begin{eqnarray}
F_{H}^{\text{in}}= \dfrac{8 \omega k}{|C|^{2}}\left|1+ K_{lm\omega} \dfrac{{}_{SN}C^{\rm ref}}{{}_{SN}C^{\rm trans}_{lm}}\right|^{2}, \\
F_{H}^{\text{out}}= \dfrac{8 \omega k}{|b_{0}|^{2}}\left| K_{lm\omega} \dfrac{{}_{SN}C^{\rm inc}}{{}_{SN}C^{\rm trans}}\right|^{2},
\end{eqnarray}
therefore justifying our requirement (\ref{requirement}).

\bibliography{apssamp}

\end{document}